\documentclass[12pt,a4paper]{article}
\usepackage{amsmath,amssymb,hyperref,bm,bbm}
\usepackage[utf8]{inputenc}
\usepackage{graphicx}
\numberwithin{equation}{section}
\allowdisplaybreaks[3]

\begin{document}
\begin{titlepage}
		
\renewcommand{\thefootnote}{\fnsymbol{footnote}}
\begin{flushright}
\begin{tabular}{l}
YITP-19-49\\
\end{tabular}
\end{flushright}
		
\vfill
\begin{center}
			
			
\noindent{\large \textbf{Rectangular W-(super)algebras and their representations}}
			

			
\vspace{1.5cm}

\noindent{Thomas Creutzig$^{a}$\footnote{E-mail: creutzig@ualberta.ca} 
	and Yasuaki Hikida$^b$\footnote{E-mail: yhikida@yukawa.kyoto-u.ac.jp}}
\bigskip

\vskip .6 truecm
\centerline{\it $^a$Department of Mathematical and Statistical Sciences, University of Alberta,} \centerline{\it Edmonton, Alberta T6G 2G1, Canada}
\medskip
\centerline{\it $^b$Center for Gravitational Physics, Yukawa Institute for Theoretical Physics,}
\centerline{\it  Kyoto University, Kyoto 606-8502, Japan}

\end{center}
		
\vfill
\vskip 0.5 truecm

\begin{abstract}

We study W-algebras obtained by quantum Hamiltonian reduction of $sl(Mn)$ associated to the $sl(2)$ embedding of rectangular type. The algebra can be realized as the asymptotic symmetry of higher spin gravity with $M \times M$ matrix valued fields. In our previous work, we examined the basic properties of the W-algebra  and claimed that the algebra can be realized as the symmetry of Grassmannian-like coset even with finite central charge based on a proposal of holography. In this paper, we extend the analysis in the following ways. First, we compute the operator product expansions among low spin generators  removing the restriction of $n = 2$. Second, we investigate the degenerate representations in several ways, and see the relations to the coset spectrum and the conical defect geometry of the higher spin gravity. For these analyses, we mainly set $M=n=2$. Finally, we extend the previous analysis by introducing $\mathcal{N}=2$ supersymmetry.

\end{abstract}
\vfill
\vskip 0.5 truecm
		
\setcounter{footnote}{0}
\renewcommand{\thefootnote}{\arabic{footnote}}
\end{titlepage}
	
\newpage
	
\tableofcontents

\section{Introduction}

Three dimensional gravity on anti-de Sitter (AdS) space is  an attractive arena to investigate the quantum aspects of gravity because of its tractability. In particular, there are infinite dimensional symmetries near the AdS boundary \cite{Brown:1986nw}, and the quantum gravity effects should be constrained by these symmetries.
In our previous work \cite{Creutzig:2018pts}, we investigated the rectangular W-algebra with $su(M)$ symmetry as the asymptotic symmetry of higher spin gravity with $M \times M$ matrix valued fields constructed in \cite{Prokushkin:1998bq}. 
The matrix extension is expected to be useful to see stringy effects from the viewpoints of higher spin gravity, see, e.g., \cite{Vasiliev:2018zer} for recent arguments. 
In \cite{Creutzig:2013tja}, it was proposed that the matrix extension of classical higher spin gravity is dual to the coset 
\begin{align}
\frac{su(N+M)_k}{su(N)_k \oplus u(1)_{\kappa}}  \label{coset}
\end{align}
with $\kappa = k N M (N + M)$ at a large $N$ limit.%
\footnote{See \cite{Candu:2013fta,Eberhardt:2018plx,Kumar:2018dso} for related works on the holography with ``stringy'' cosets.}
For $M=1$, the proposal reduces to the Gaberdiel-Gopakumar duality \cite{Gaberdiel:2010pz}.
Based on the holography of \cite{Creutzig:2013tja}, we claimed that the rectangular W-algebra can be realized as the symmetry algebra of \eqref{coset} even without taking a large $N$ limit.
In this paper, we continue the study of the rectangular W-algebra by extending the analysis of operator product expansions (OPEs) among generators and examining its degenerate representations. 
We further examine the $\mathcal{N}=2$ supersymmetric extensions of rectangular W-algebras.

It is known that pure AdS gravity in three dimensions can be described by $sl(2) \oplus sl(2)$ Chern-Simons gauge theory \cite{Achucarro:1987vz,Witten:1988hc}, and a higher spin gravity can be constructed by replacing $sl(2)$ by a higher rank gauge algebra, say, $sl(n)$.
In order to realize a matrix extension, we may consider  the gauge algebra with the multiplication of $M \times M$ matrix algebra or $gl(M)$. A closed algebra including $gl(M) \otimes sl(n)$ is given by
(see, e.g., \cite{Gaberdiel:2013vva,Creutzig:2013tja,Joung:2017hsi})
\begin{align}
sl(M n) \simeq sl(M) \otimes \mathbbm{1}_n \oplus \mathbbm{1}_M \otimes sl(n) \oplus sl(M) \otimes sl(n) \, , \label{slMn}
\end{align}
and the gravitational sector is identified with the principally embedded $sl(2)$ in $\mathbbm{1}_M \otimes sl(n)$.  The asymptotic symmetry is obtained as in \cite{Campoleoni:2010zq,Henneaux:2010xg} by assigning the asymptotic AdS condition, and it is identified as a W-algebra given by the Hamiltonian reduction of $sl(M n)$ with the corresponding $sl(2)$ embedding. 
In particular, the W-algebra includes the Virasoro algebra and $su(M)$ affine  algebra as subalgebras.
In \cite{Creutzig:2018pts}, we computed the exact expressions of the central charge $c$ and the level $\ell$ of $su(M)$ currents. We further obtained the OPEs among generators for $n=2$ by requiring their associativity. 
We claimed that the rectangular W-algebra with $su(M)_k$ symmetry is realized by the coset \eqref{coset} at $\lambda =n$.
Here the 't Hooft parameter $\lambda$ is defined by
\begin{align}
\lambda = \frac{k}{k + N} \quad \text{or} \quad  \lambda = - \frac{k}{k+N+M} \, , \label{tHooft}
\end{align}
which can be exchanged by the duality of cosets \cite{Creutzig:2018pts}.
We have checked that the OPEs among generators are reproduced from the coset at $\lambda = 2$.

In this paper, we extend the analysis in the following ways.
We first re-examine the OPEs of low spin generators by working with generic $n$.
In our previous analysis, we assumed the decoupling of spin 3 currents at $n=2$ and obtained OPEs among generators up to spin 2. Here we remove the assumption and examine the OPEs between spin 2 and 3 currents as well. Since the computations become rather complicated, we set $M=2$ for simplicity.
We then analyze the degenerate representations of the rectangular W-algebra in several ways mainly working with $M=n=2$. As a direct way, we study the conditions for null vectors at the low levels of descendants. We find several examples of them by assuming a certain form of null vectors. We further obtain representations from those of $sl(Mn)$ current algebra by applying the Hamiltonian reduction.  We compare the results with the spectrum of the coset \eqref{coset} and the mass of conical defect geometry in the Chern-Simons theory obtained in \cite{Castro:2011iw}, see also \cite{Gaberdiel:2012ku,Perlmutter:2012ds,Hikida:2012eu}.

We further extend the analysis by considering the $\mathcal{N}=2$ supersymmetric rectangular W-algebras.
It is important to introduce  extended supersymmetry in order to see the relation to superstring theory \cite{Creutzig:2011fe,Gaberdiel:2013vva,Creutzig:2013tja,Creutzig:2014ula,Gaberdiel:2014cha}.
The $\mathcal{N}=2$ W-algebra may be realized as the asymptotic symmetry of  $\mathcal{N}=2$ higher spin supergravity with $M \times M$ matrix valued fields in \cite{Prokushkin:1998bq}. We construct the $\mathcal{N}=2$ W-algebra by the Hamiltonian reduction of $sl(M (n+1)| M n)$.
The $\mathcal{N}=2$ W-algebra includes the Virasoro algebra and two $su(M)$ affine algebras as subalgebras.
We compute the central charge $c$ and the levels $\ell_1,\ell_2$ of two $su(M)$ currents without taking a large $c$ limit.
We also obtain the OPEs among low spin generators for simple examples with $n=1$ and $M=2,3,4$.
Based on the holography of \cite{Creutzig:2013tja}, we claim that the $\mathcal{N}=2$ W-algebra is realized as the symmetry of the coset
\begin{align}
\frac{su(N+M)_k \oplus so(2 NM)_1}{su(N)_{k+M} \oplus u(1)_{\hat \kappa}}   \label{scoset}
\end{align}
with $\hat \kappa = N M (N + M) (N + M + k)$ even at finite $c$.
From the match of central charges and the levels of $su(M)$ currents, we set
\begin{align}
\ell_1 = k \, , \quad \ell_2 = N \, , \quad \lambda \equiv \frac{N}{k+N+M} = - n \, , \label{smap1}
\end{align}
or 
\begin{align}
\ell_1 = N \, , \quad \ell_2 = k \, , \quad \lambda \equiv \frac{k}{k + N +M} = - n \, . \label{smap2}
\end{align}
We check that the OPEs among generators can be reproduced from the coset \eqref{scoset} for several examples.%
\footnote{It is known that the coset \eqref{scoset} with $M=2$ has the large $\mathcal{N}=4$ superconformal symmetry, and a holography with the coset has been proposed in \cite{Gaberdiel:2013vva}.
In the special case with $M=2$, the symmetry generators of the coset for low spins were explicitly constructed in \cite{Ahn:2013oya,Gaberdiel:2014yla}, and the OPEs among generators of the $\mathcal{N}=4$ W-algebra were investigated in \cite{Beccaria:2014jra}.}

This paper is organized as follows.
In the next section, we start by reviewing the results of \cite{Creutzig:2018pts}.
We explain the derivations of the central charge and the level of $su(M)$ currents.
We then compute the OPEs among generators but now with generic $n \neq 2$ but with $M=2$. 
We compare these results with those of the dual coset \eqref{coset}. In section \ref{sec:degenerate}, we examine the degenerate representations from the null vector conditions and the Hamiltonian reduction of $sl(4)$.  We then compare the results with the spectrum of dual coset \eqref{coset} and the mass of conical defect geometry of higher spin gravity.
In section \ref{sec:N=2}, we examine the $\mathcal{N}=2$ rectangular W-algebra.
We first derive the expressions of central charge and the levels of two $su(M)$ currents.
We then compute the OPEs among generators of spin up to 2 for small $M$ and examine degenerate representations. We compare these results with those of dual coset \eqref{scoset}.
Section \ref{sec:conclusion} is devoted to conclusion and discussions.
Several technical appendices follow.
In appendix \ref{sec:CR}, we write down the commutation relations for the bosonic W-algebra with $M = n=2$. In appendix \ref{sec:minimal}, the degenerate representations of the bosonic W-algebra are re-examined by making use of associativity of the OPEs. In appendix \ref{sec:N=2info}, we collect several technical materials used for the analysis of the $\mathcal{N}=2$ W-algebra. In appendix \ref{sec:alternative}, we propose alternative coset realizations of the rectangular W-(super)algebras.

\section{Rectangular W-algebra}

As mentioned in the introduction, a higher spin gravity can be constructed by the Chern-Simons gauge theory based on a higher rank algebra $\mathfrak{g}$ with an embedding of gravitational $sl(2)$.
Without the matrix extension, a holography was proposed in \cite{Gaberdiel:2010pz} using the 3d Prokushkin-Vasiliev theory of \cite{Prokushkin:1998bq}. The gauge algebra of the 3d higher spin theory is given by $hs[\lambda]$, which can be truncated to $sl(n)$ at $\lambda = n$. Similarly, the gauge algebra of higher spin theory with the matrix extension is given by $hs_M[\lambda]$, which can be reduced to $sl(M n)$ at $\lambda =n$.
Decomposing $sl(M n)$  as in \eqref{slMn}, we principally embed the gravitational $sl(2)$ in $\mathbbm{1}_M \otimes sl(n)$. 
The gauge algebra can be decomposed by the $sl(2)$ as
\begin{align}
sl(Mn) \simeq sl(M) \oplus M^2 g^{(2)} \oplus \cdots \oplus M^2 g^{(n)} \, ,
\label{slMndec}
\end{align}
where $g^{(s)}$ denotes the spin $s-1$ representation of $sl(2)$.
After the Hamiltonian reduction, only one element in $g^{(s)}$ $(s=2,3,\ldots,n)$ survives and the space-time spin of the element becomes $s$.
Thus, the W-algebra includes $sl(M)$ (or $su(M)$) spin 1 currents%
\footnote{We consider the complex elements of $sl(Mn)$, and thus there is no distinction between $sl(M)$ and $su(M)$.} 
and $M^2$ spin $s$ currents with $s=2,3,\ldots , n$. 
In particular, one of  the spin 2 currents is the energy-momentum tensor.

In this section, we explain the basic properties of rectangular W-algebras by reviewing the results of \cite{Creutzig:2018pts}. In the next subsection, we compute the central charge $c$ and the level $\ell$ of the $su(M)$ currents by applying the general formula, e.g., in \cite{Kac:2003jh}.
In subsection \ref{sec:OPE}, we obtain the OPEs among generators by requiring their associativity.  
Here we do not assume the decoupling of spin 3 currents  but work with $M=2$ just for simplicity.
In subsection \ref{sec:dualcoset}, we examine the relations to the dual coset of \eqref{coset}.

\subsection{Central charge and level of the affine symmetry}
\label{sec:basic}

In \cite{Creutzig:2018pts}, we computed the central charge $c$ and the level $\ell$ of $sl(M)$ current
as basic information on the W-algebra with $sl(M)$ symmetry.
Here we repeat the analysis mainly for the preparation of supersymmetric extension.
We start with a Lie (super)algebra $\mathfrak{g}$ and set the level of the affine currents as $t$.
We also specify an embedding of $sl(2)$.
We may denote the generators of $sl(2)$ by $x,e,f$ satisfying
\begin{align}
[x ,e] = e \, , \quad [x,f] = - f \, , \quad  [e , f] = x \, .
\label{sl2}
\end{align}
The Lie (super)algebra can be decomposed by the eigenvalue of the adjoint action $\text{ad} \, x$ as
\begin{align}
\mathfrak{g} = \oplus_{j \in \frac12 \mathbb{Z}} \mathfrak{g}_j \, .
\end{align}
Denoting $S_+ = \prod_{j > 0} S_j$ with a basis $\{ u_\alpha \}_{\alpha \in S_j}$ for $\mathfrak{g}_j$, the formula for the central charge is \cite{Kac:2003jh} 
\begin{align}
\begin{aligned}
c =& \frac{t \, \text{sdim} \, \mathfrak{g}}{t + h^\vee} - 12 t (x|x) 
 - \sum_{\alpha \in S_+} (-1)^{p (\alpha)} (12 m^2_\alpha  - 12 m_\alpha + 2) - \frac12 \, \text{sdim} \,  \mathfrak{g}_{1/2} \, .
\end{aligned}
\label{KW}
\end{align}
Here $(x|x)$ is a bilinear form and $h^\vee$ represents the dual Coxeter number.
The parity of $u_\alpha$ is denoted by $p(\alpha)$.
The Hamiltonian reduction is realized as a BRST cohomology at the quantum level and the conformal dimensions of BRST ghosts are $(m_\alpha, 1 - m_\alpha)$ with $m_\alpha = 1 - j$.

For the present case, we set $\mathfrak{g}=sl(Mn)$, where the dimension is $M^2 n^2 - 1$ and the dual Coxeter number is $h^\vee = M n$.
With $x$ for the principally embedded $sl(2)$ in $\mathbbm{1}_M \otimes sl(n)$, we find 
\begin{align}
(x|x) =  \text{tr} (x x) = \text{tr} \mathbbm{1}_M \cdot \frac{1}{12} n (n^2 -1)   = \frac{1}{12} Mn (n^2 -1) \, .
\end{align}
We next count the number of elements belonging to $S_j$, which is found to be $M^2 (n-j)$ for $j=1,2,\ldots, n-1$ and zero otherwise. 
From the formula \eqref{KW}, the total central charge is obtained as
\begin{align}
\begin{aligned}
c = \frac{t (n^2 M^2 -1)}{t + nM} - t M n (n^2 -1)  - M^2 n (1 + (-2 + n) n^2) \, . \label{DSc}
\end{aligned}
\end{align}

We compute the level $\ell$ of $sl(M)$ currents as well.
The currents basically come from the elements in $sl(M) \otimes \mathbbm{1}_n$, which leads to the level $t n$.
There are also contributions from the BRST ghosts.
The ghosts carry the trivial and adjoint representations of $sl(M)$, and there are $2 \sum_{j=1}^{n-1} (n-j) = n (n-1) $ sets of ghosts in the adjoint representation. We can construct $sl(M)$ currents with level $M$ from one set of ghosts in the adjoint representation. Therefore, the sum of two types of contributions to the level is
\begin{align}
\ell =  t n + M n (n -1) \, .  \label{DSl}
\end{align}
We have a relation between $c$ and $\ell$ as
\begin{align}
c = - \frac{(\ell ^2 - 1) n^2 M}{ \ell + n M} + \ell M - 1 
\label{clrel}
\end{align}
by expressing $t$ in terms of $\ell$.

\subsection{OPEs among low spin currents}
\label{sec:OPE}

We can examine the possible OPEs among generators of algebra by requiring their associativity.%
\footnote{For the checks of associativity of OPEs, we use a Mathematica package ``OPEdefs''  \cite{Thielemans:1991uw}.
}
In \cite{Creutzig:2018pts}, the OPEs among generators of the rectangular W-algebra were obtained at $n=2$ by assuming the decoupling of spin 3 currents. Here we lift the analysis to generic $n$.

As seen above, the rectangular W-algebra includes $su(M)$ affine algebra as a sub-algebra.
We denote the currents as $J^a$ with $a=1,2,\ldots , M^2- 1$ and require the OPEs
\begin{align}
J^a (z) J^b(0) \sim \frac{\ell \delta^{ab}}{z^2} + \frac{i f^{ab}_{~~c} J^c (0)}{z} 
\label{OPEJJ}
\end{align}
with the level $\ell$.
Here we have introduced the generators $t^a$ of $su(M)$ with the invariant tensors
\begin{align}
\text{tr} (t^a t^b) = \delta^{ab} \, , \quad 
\text{tr}( [t^a , t^b] t^c ) = i f^{abc} \, , \quad
\text{tr} (\{ t^a , t^b \} t^c) = d^{abc} \, . \label{invtensors}
\end{align}
The algebra also includes higher spin currents $W^{(s)A}$ $(s=2,3,\ldots)$, where $W^{(s)0}$ and $W^{(s)a}$ are in the trivial and adjoint representations of $su(M)$. In particular, $T \equiv W^{(2)0}$ is the energy-momentum tensor satisfying
\begin{align}
T (z) T(0) \sim \frac{c/2}{z^4} + \frac{2 T(0)}{z^2} + \frac{T ' (0)}{z} \, , \quad
T (z) J^a (0) \sim \frac{J^a (0)}{z^2} + \frac{{J^a}' (0)}{z} 
\label{OPETJ}
\end{align}
with the central charge $c$.
Here $A' (0)$ means $\frac{d}{dz} A(z)$ at $z = 0$.
We choose the primary basis for the charged higher spin currents $Q^a \equiv W^{(2)a}$ and $P^a \equiv W^{(3)a}$ such as to satisfy
\begin{align}
\begin{aligned}
T (z) Q^a (0) \sim \frac{2 Q^a (0)}{z^2} + \frac{Q^a {}' (0)}{z} \, , \quad
J^a (z) Q^b (0) \sim \frac{i f^{a b}_{~~c} Q^c (0)}{z} \, , \\
T (z) P^a (0) \sim \frac{3 P^a (0)}{z^2} + \frac{P^a {}'(0)}{z} \, , \quad
J^a (z) P^b (0) \sim \frac{i f^{a b}_{~~c} P^c (0)}{z} \, .
\end{aligned}
\label{OPEQP}
\end{align}
The first non-trivial OPE would be among $Q^a$ and $Q^b$, which will be expressed schematically as $Q^a \times Q^b$.

\subsubsection{Composite primary operators}

The operator product $W^{(s_1)A_1} \times W^{(s_2)A_2}$ produces (composite) operators up to spin $s_1 +s_2 -1$. For the OPE of $Q^a \times Q^b$, we need (composite) operators with spin 1,2,3, though we can see that $W^{(3)0}$ does not appear  due to the bosonic statistic of $Q^a$.
We restrict the form of $Q^a \times Q^b$ by requiring the associativity of  $Q^a \times Q^b \times Q^c$,
but for this we also need to think about the OPEs of $Q^a$ and $P^b$. The operator product $Q^a \times P^b$ would generate composite primary operators of spin up to 4, so we start by classifying them. 

We may count the number of primary operators by decomposing the vacuum character of 
the rectangular W-algebra by the Virasoro characters, see, e.g., \cite{Candu:2012tr}.
The vacuum character of the rectangular W-algebra
is
\begin{align}
\chi^M_\infty (q) =  \prod_{n=1}^\infty \frac{1}{(1 - q^n)^{M^2 -1}}\prod_{s=2}^\infty \prod_{n=s}^\infty \frac{1}{(1 - q^n)^{M^2}} \, ,
\end{align}
while the Virasoro characters of the vacuum representation and the generic one with conformal weight $h$ are
\begin{align}
\chi_0 (q) = \prod_{n=2}^\infty \frac{1}{1 - q^n} \, , \quad \chi_h (q) = \frac{q^h}{1 - q} \chi_0 (q) \, .
\label{Virch}
\end{align}
The decomposition is given by
\begin{align}
\chi_\infty^M (q) = \chi_0 (q) + \sum_{h = 1}^\infty d(h) \chi_h (q) \, ,
\end{align}
where $d(h)$ represents the number of independent (composite) primaries with conformal weight $h$.
Expanding in $q$, we find
\begin{align}
&d(1) = M^2 -1 \, , \quad d(2) = \frac{1}{2} \left(M^4+M^2-2\right) \, , \quad d(3) = \frac{1}{6} \left(M^6+9 M^4-16 M^2+12\right) \, , \nonumber \\
&d(4) = \frac{1}{24} \left(M^8+22 M^6+23 M^4-46 M^2+24\right) 
\label{dn}
\end{align}
and so on.

We can construct composite operators primary w.r.t.~the Virasoro generator along with the fundamental currents $J^a,W^{(s)A}$ with $s=2,3,\ldots$. 
In order to define the composite operators, we adopt the prescription of normal ordering as
\begin{align}
(A B) (z) =  \frac{1}{ 2 \pi i } \oint \frac{d w}{w - z} A(w) B(z) 
\label{NO}
\end{align}
and 
\begin{align}
 (A_1  \cdots A_{l-2} A_{l-1} A_l) = (A_1  \cdots ( A_{l-2} (A_{l-1} A_l)) \cdots  ) \, .
\end{align}
Moreover, we use the brackets $(a_1 , \ldots , a_l)$ and $[a_1, \ldots , a_l]$ for the symmetric and anti-symmetric indices with pre-factor $1/(l!)$, respectively.
We find the composite operators
\begin{align}
[J^{(a} J^{b)}] = (J^{(a} J^{b)})  -  \frac{2 \ell }{c} \delta^{ab} T
\end{align}
for spin 2 and
\begin{align}
\begin{aligned}
&[ J^{(a)} J^{b} J^{c)} ] = (J^{(a)} J^{b} J^{c)}) + \frac{3 \ell }{c+2} \left[ \delta^{(ab}J^{c)}{ }''-2 (T J^{(a})  \delta^{bc)}\right] \, , \\
&[J^a Q^b] = (J^a Q^b)  - \frac{i}{4} f^{ab}_{~~c}  Q^{c} {} ' \, , \\
&[J^{[a }  J^{b]}{ } '] = (J^{[a }  J^{b]} {} ') - \frac{i}{3} f^{ab}_{~~c} \left[ \frac{c + 5}{c + 2 } J^c{} '' - \frac{6}{c + 2 } (T  J^c) \right] 
\end{aligned}
\end{align}
for spin 3.
For general composite operators, we use the abbreviated notation
\begin{align}
[A_1 \cdots  A_l ] = ( A_1 \cdots  A_l) + \cdots \, , \label{ab}
\end{align}
where the dots represent terms which make the operators to be primary w.r.t.~the Virasoro algebra.%
\footnote{The primary condition does not fix the additional terms uniquely. However, we do not write down our specific choices here, since they are not so important for our arguments.}
With the notation, the composite spin 4 primary operators are
\begin{align}
\begin{aligned}
&[ J^{(a} J^{b} J^{c} J^{d)} ] \, , \quad [J^{(a } J^{b)} Q^c] \, , \quad
[J^{(a} J^{b)}{}'']  \, , \quad
[J^a  Q^b {} ']  \, , \\
&[Q^{(a} Q^{b)}]  \, , \quad
( J^a W^{(3)0}) \, , \quad [J^a P^b]  
\end{aligned}
\label{CPSpin4}
\end{align}
and
\begin{align}
 \left[ J^{(a } J^{b)} { J^c} ' \right]   = \frac14 \{ (J^a J^b {J^c}') + (J^b J^a {J^c}') - (J^c J^b {J^a}') - (J^c J^a {J^b}') \} + \cdots \, . 
\end{align}
We can check that the number of primary operators matches with $d(h)$ in \eqref{dn} up to $h=4$.
Notice that the number of independent components of $[J^{(a} J^{b)} J^{c}{}']$ is $L (L^2 -1)/3$ with $L = M^2 -1$.

\subsubsection{ Associativity of OPE}

We expand the operator product $Q^a \times Q^b$ in terms of the  fundamental and composite operators with some indices. The coefficients in front of the operators can be expressed with the invariant tensors of $su(M)$.
Here we restrict ourselves to the case with $M=2$, but it is straightforward (but tedious) to work with generic $M$. There is only one tensor with 2 indices, that is $\delta^{ab}$. With 3 indices, there are $f^{abc}$ and $d^{abc}$ introduced in \eqref{invtensors}. For $M=2$, we have $d^{abc} = 0$, and the other invariant tensors can be written in terms of $\delta^{ab}$ and $f^{abc}$. With 4 indices, there are
\begin{align}
d_{4AA1}^{abcd} = \frac12 (  \delta^{ac} \delta^{bd} - \delta^{ad} \delta^{bc} ) \, , \quad
d_{4SS1}^{abcd} = \delta^{ab} \delta^{cd} \, , \quad 
d_{4SS2} =  \frac12 (  \delta^{ac} \delta^{bd} + \delta^{ad} \delta^{bc} ) \, . \label{d4}
\end{align}
Furthermore, we need
\begin{align}
\begin{aligned}
&d_{5AS1}^{abcde} = i f^{ab(c} \delta^{de)} \, , \quad
d_{5AH1}^{abcde} = \frac{i}{4} (2 f^{abe} \delta^{cd} - f^{abc} \delta^{ed} - f^{abd} \delta^{ec}) \, , 
 \\
&d_{5SH}^{abcde} = \frac{i}{4} (\delta^{ac} f^{bde} + \delta^{bc}f^{ade} 
+ \delta^{ad}f^{bce} + \delta^{bd}f^{ace})
\end{aligned} \label{d5}
\end{align}
for those with 5 indices and
\begin{align}
d_{6SS1}^{abcdef} =  \delta^{ab} \delta^{c(d} \delta^{ef)} \, , \quad
d_{6SS2}^{abcdef} = \frac{1}{6} d^{abcd}_{4SS2} \delta^{ef} + \cdots  \label{d6}
\end{align}
for those with 6 indices. Here the dots above are the terms which make the expression symmetric under the exchange of $\{c,d,e,f\}$.

With the above preparations, we require the following forms of the OPEs involving $Q^a$ and $P^a$.
The OPE of $Q^a \times Q^b$ is schematically of the form 
\begin{align}
\begin{aligned}
Q^a \times Q^b \sim \, &\delta^{ab} c_1/2 \cdot   \mathbbm{1} + i f^{ab}_{~~c} c_2 J^c + C^{ab}_{3,cd} [J^{(c } J^{d)}] + d_{4AA1,cd}^{ab} c_4 [J^{[c} J^{d]} {}']  \\ & + d_{5AS1,cde}^{ab} c_5 [J^{(c} J^d J^{e)}] + d_{4AA1,cd}^{ab} c_6 [J^c Q^d] + i f^{ab}_{~~c} c_7 P^c \, .
\end{aligned}
\end{align}
There are also contributions from descendants, which can be related to those from the primaries utilizing the Virasoro symmetry or requiring the associativity of $T \times Q^a \times Q^b$.
Here and in the following, we use small letters like $c_i$ for constants without indices and capital ones like $C_{i}$  for coefficients with indices. In the current example, we set 
\begin{align}
C^{ab}_{3,cd} = d_{4SS1,cd}^{ab} c_{31} + d_{4SS2,cd}^{ab} c_{32} \, .
\label{C3abcd}
\end{align}
Requiring the associativity of $J^a \times Q^b \times Q^c$, almost all coefficients are fixed as functions of $c$ and $\ell$ up to 2 parameters.
One of them may be chosen as $c_1$, which can be absorbed by changing the overall normalization of $Q^a$.
We choose the other parameter as $c_7$, which is undetermined at this stage.

The OPE of $Q^a \times P^b$ is of the form
\begin{align}
Q^a \times P^b &\sim i f^{ab}_{~~c} e_1 J^c + E^{ab}_{2,cd} [J^{(c } J^{d)}] + i f^{ab}_{~~c} e_3 Q^c + d^{ab}_{5AS,cde} e_4 [J^{(c} J^d J^{e)}] + E^{ab}_{5,cd} [J^c Q^d] \nonumber  \\
&+ d^{ab}_{4AA,cd} e_6  [J^{[c} J^{d]} {}'] + \delta^{ab} e_7 W^{(3)0} + i f^{ab}_{~~c} e_8 P^c + E^{ab}_{9,cdef} [J^{(c} J^d J^e J^{f)}]
\\
& + E^{ab}_{10,cde}[J^{(c} J^{d)} J^e {}'] + E^{ab}_{11,cd} [J^{(c} J^{d)} {}''] + E^{ab}_{12,cde}[J^{(c} J^{d)} Q^e] + E^{ab}_{13,cd} [J^c Q^d {}'] \nonumber \\
&+ E^{ab}_{14,cd} [Q^{(c} Q^{d)}]  + i f^{ab}_{~~c} e_{15} (J^c W^{(3)0}) + E^{ab}_{16,cd} [J^c P^d] + \delta^{ab} e_{17}  W^{(4)0} + i f^{ab}_{~~c} e_{18} W^{(4)c} \, . \nonumber 
\end{align}
The coefficients with capital letters as $E_{i}$ are expressed by the invariant tensors in \eqref{d4}, \eqref{d5} and \eqref{d6} as in \eqref{C3abcd}.
This OPE is used only to restrict the form of $Q^a \times Q^b$ from the associativity of $Q^a \times Q^b \times Q^c$.
For this, we do not need the information of $e_7, e_{15},E_{16}, e_{17},e_{18}$, which will be neglected in the following. Requiring the associativity of $J^a \times Q^b \times P^c$, the other coefficients are determined as functions of $c$ and $\ell$ up to 3 parameters including $c_1,c_7$. One remaining parameter may be chosen as $e_3$, which can be absorbed by changing the overall normalization of $P^a$.

The associativity of $Q^a \times Q^b \times Q^c$ leads to a constraint equation for these parameters as
\begin{align}
\begin{aligned}
e_3 c_7 \ell \left(3 \ell^2+\ell-4\right) (c \ell +2 c-3 \ell)
= c_1 (\ell-2) (\ell+1) \left(c \ell +4 c+6 \ell ^2-7 \ell -4\right) \, .
\end{aligned}
\label{constraint}
\end{align}
Since $c_1$ and $e_3$ can be removed by the redefinitions of $Q^a$ and $P^a$, we conclude that the OPEs of generators of spin up to $2$ have only two parameters $c$ and $\ell$. 
The decoupling of spin 3 currents $P^a$ can be realized with $c_7 = 0$, which leads to
\begin{align}
c = - \frac{8 (\ell^2 -1)}{\ell + 4} + 2 \ell - 1 
\end{align}
as in \eqref{clrel} with $M=n=2$.
In terms of mode expansions, the commutation relations among generators  with $M=n=2$ are summarized in appendix \ref{sec:CR}.

From the above analysis, we may conjecture that there is a W-algebra with $su(M)$ symmetry parametrized by $c$ and $\ell$. Based on the relation in \eqref{clrel} with $M=2$, we may map a parameter $c$ to $\lambda$ by
\begin{align}
c = - \frac{2 (\ell^2 -1) \lambda^2 }{ \ell + 2 \lambda} + 2 \ell - 1 \, .
\end{align}
The parameters of the W-algebra are now $\lambda$ and $\ell$, and the W-algebra from $sl(2n)$ is expected to be realized at $\lambda = n$.
Except for $\ell = \pm 1 , - 2 \lambda$, the constraint equation \eqref{constraint} reduces to
\begin{align}
\begin{aligned}
 &e_3 c_ 7  \ell  \left(3 \ell ^2+\ell -4\right) (\lambda-1) ((\ell +2) \lambda+ \ell) \\
 & \qquad = c_1 (\ell -2) (\ell +1) (\lambda-2) (\ell  (\lambda+2)+4 \lambda)  \, .
\end{aligned}
\label{constraint2}
\end{align}
This is a second order equation with respect to $\lambda$. Suppose that $\lambda = \lambda_0$ is a solution to \eqref{constraint2}, then we can show that
\begin{align}
\lambda = - \frac{\ell  \lambda_0}{ \ell + 2 \lambda_0 }
\label{dual0}
\end{align}
 is also a solution.
 As seen shortly, this is consistent with the duality relation discussed in \cite{Creutzig:2018pts}.

\subsection{Dual coset CFT}
\label{sec:dualcoset}

In \cite{Creutzig:2013tja}, it was proposed that the classical 3d Prokushkin-Vasiliev theory of \cite{Prokushkin:1998bq} with $M \times M$ matrix valued fields is dual to the Grassmannian-like coset \eqref{coset} at a large $N$ limit.
We have checked the agreements of spectrum and low spin symmetry in the limit. In \cite{Creutzig:2018pts}, we conjectured that the holographic duality works even with finite $N$ and claimed that the rectangular W-algebra with $su(M)$ symmetry can be realized as the symmetry algebra of the coset \eqref{coset}.%
\footnote{An alternative coset description of the same W-algebra is proposed in appendix \ref{sec:alternativeb} and several confirmations are given. It is an interesting open problem to examine more direct relations between the two dual cosets.}
The correspondence happens at $ \lambda = n $, where the 't Hooft parameter is defined by \eqref{tHooft}.
We should also set $\ell = k$, where $\ell$ is the level of $su(M)$ current algebra.
The central charge is written as 
\begin{align}
c = - \frac{(k^2 - 1) \lambda^2 M}{ k + \lambda M} + k M  - 1 
\label{centerlambda}
\end{align}
irrespective of the choice of the 't Hooft parameter \eqref{tHooft}.
Notice that the expression with $\lambda = n$ reduces to \eqref{clrel}.
The two choices are related to a duality of the coset \eqref{coset} as discussed in \cite{Creutzig:2018pts}.
In particular, the two 't Hooft parameters can be exchanged by
\begin{align}
\lambda \leftrightarrow - \frac{\lambda k}{k + \lambda M} \, ,
\end{align}
which is consistent with \eqref{dual0}.

The generators of the rectangular W-algebra have been constructed in terms of the coset \eqref{coset} in \cite{Creutzig:2018pts} up to spin 3.
Here we review the results of \cite{Creutzig:2018pts} in order to prepare for the supersymmetric extension.
We decompose $su(N+M)$ as
\begin{align}
su(N + M) = su(N) \oplus su(M) \oplus u(1) \oplus (N , \bar M) \oplus (\bar N , M)  \label{suNMdec}
\end{align}
and use the generators $t^A = (t^\alpha , t^a , t^{u(1)} , t^{( \rho \bar \imath ) } , t^{ (  \bar \rho  i ) })$.
Here $L$ and $\bar L$ represents the fundamental and anti-fundamental representations of $su(L)$, respectively.  
We introduce the metric $g^{AB} = \text{tr} (t^A t^B)$ as
\begin{align}
\text{tr} (t^\alpha t^\beta) = \delta^{\alpha \beta} \, , \quad
\text{tr} (t^a t^b) = \delta^{ab} \, , \quad
\text{tr} (t^{u(1)} t^{u(1)}) = 1 \, , \quad
\text{tr} (t^{(\rho \bar \imath)} t^{(\bar \rho i)}) =  \delta^{\rho \bar \rho}\delta^{i \bar \imath} 
\label{suNMmetric}
\end{align}
and the invariant tensors as
\begin{align}
i f^{ABC} = \text{tr} ([t^A ,t^B] t^C)  \, , \quad d^{ABC} = \text{tr} (\{ t^A , t^B\} t^c)   \, .
\label{suNMtensor}
\end{align}
We adopt the convention in \cite{Creutzig:2014ula,Creutzig:2018pts} such that
\begin{align}
\begin{aligned}
&i f^{(\rho \bar \imath) (\bar \sigma j) u(1)} = \sqrt{\frac{M+N}{MN}} \delta^{\bar \imath j  } \delta^{\bar \sigma \rho } \, , \quad
d^{(\rho \bar \imath) (\bar \sigma j) u(1)} = \frac{M-N}{\sqrt{MN(N+M)}} \delta^{ \bar \imath  j } \delta^{\bar \sigma \rho } \, , \\
&i f^{(\rho \bar \imath) (\bar \sigma j) \alpha} = d^{(\rho \bar \imath) (\bar \sigma j) \alpha} =  (t^\alpha)^{\bar \sigma \rho}  \delta^{j \bar \imath }  \, , \quad
i f^{(\rho \bar \imath) (\bar \sigma j) a} = - d^{(\rho \bar \imath) (\bar \sigma j) a} = - \delta^{\rho \bar \sigma} (t^a)^{\bar \imath j}  
\end{aligned} \label{exptensor}
\end{align}
for non-trivial expressions.
With these notations, the $su(M)$ currents $J^A$ satisfy
\begin{align}
J^A (z) J^B (0) \sim \frac{k g^{AB}}{z^2} + \frac{i f ^{AB}_{~~~C} J^C ( 0)}{z} \, .
\label{suNMJJOPE}
\end{align}
The spin 1 current is given by $J^a$ from $su(N+M)_k$ in the numerator of \eqref{coset}.
The energy momentum tensor $T$ can be obtained from the standard coset construction \cite{Goddard:1984vk}. For the charged spin 2 currents, 
we found that
\begin{align}
\begin{aligned}
Q^a =& [(J^{(\rho \bar \imath)} J^{( \bar \rho j )}) +( J^{( \bar \rho j )}J^{(\rho \bar \imath)}) ] \delta_{\rho \bar \rho} (t^a)_{j \bar \imath} \\
& \qquad  - \frac{N}{M + 2k} d^{a}_{~bc} (J^b J^c) 
+ \frac{2}{k} \sqrt{\frac{N (N + M)}{M}} ( J^a J^{u(1)}) 
\end{aligned} \label{Qcoset}
\end{align}
satisfy the OPEs \eqref{OPEQP}.
In \cite{Creutzig:2011fe}, we have explicitly checked the match for the OPE of $Q^a \times Q^b$ at $n=2$. 
In fact, we have already shown that the associativity uniquely fixes the OPE of $Q^a \times Q^b$ with two parameters $(\lambda,\ell)$ for generic $\lambda \neq 2 $ but with $M=2$.
For the expressions of spin 3 currents, see appendix A of \cite{Creutzig:2011fe}.

\section{Degenerate representations}
\label{sec:degenerate}

In this section, we study the degenerate representations of the rectangular W-algebra in various ways.
We set $n=2$ since we know the OPEs of all generators. We furthermore restrict ourselves to the $M=2$ case just for simplicity. 
A direct way to obtain degenerate representations is to find out null states constructed from a set of primaries, which is the subject of the next subsection. We also study representations from  the Hamiltonian reduction of $sl(4)$ in subsection \ref{sec:hamilton}. 
The results are compared with the spectrum of the coset theory \eqref{coset} in subsection \ref{sec:cosetspe}
and the mass of conical defect geometry in subsection \ref{sec:conical}. 

\subsection{Null states}
\label{sec:null}

We have examined the commutation relations among generators of the W-algebras.
The next task may be to study its representations.
With $n=2$, the generators of the W-algebras are $J^a,T,Q^a$, and their mode expansions are
\begin{align}
J^a (z) = \sum_{n \in \mathbb{Z}} \frac{J^a_n}{z^{n +1}} \, , \quad T(z) = \sum_{n \in \mathbb{Z}}  \frac{L_n}{z^{n+2}} \, , \quad
Q^a (z) = \sum_{n \in \mathbb{Z}} \frac{Q^a_n}{z^{n+2}} \, . 
\end{align}
From the OPEs \eqref{OPEJJ}, \eqref{OPETJ}, and \eqref{OPEQP}, the commutation relations among these modes are obtained as
\begin{align}
\begin{aligned}
&[L_m , L_n] = (m - n) L_{m + n} + \frac{c}{12} (m^3 - m) \delta_{n+m,0} \, , \\
&[J_m^a , J_n^b] = i f^{ab}_{~~c} J^c_{m+n} + \ell m \delta_{m+n,0} \, , 
[L_m , J^a_n] = - n J^a_{m + n} \, , \\
&[L_m , Q^a_n] = (m - n) Q^a_{m+n} \, , \quad [J^a_m , Q^b_n] = i f^{ab}_{~~c} Q^c_{m+n} \, .
\end{aligned} \label{CRmodes}
\end{align} 
See  appendix \ref{sec:CR} for $[Q^a_m ,Q^b_n]$.
We look for degenerate representations since they might be used to construct the minimal models of the W-algebra. 
For instance, we will observe that several representations appear also in the spectrum of the coset \eqref{coset}  in subsection \ref{sec:cosetspe}.

For the representations of the W-algebra, we start by defining the vacuum state as
\begin{align}
L_m | 0 \rangle = 0 ~ (m \geq -1) \, , \quad Q_n^a | 0 \rangle ~ (n \geq -1) \, , \quad J^a_l | 0 \rangle~ (l \geq 0) \, .
\label{vacuumcond}
\end{align}
Notice that the vacuum is in the trivial representation of $su(2)$ since $ J^a_0 | 0 \rangle = 0$.
We further introduce states primary w.r.t.~the rectangular W-algebra by
\begin{align}
L_m | j \rangle_J = 0 ~ (m \geq 1) \, , \quad Q_n^a | j \rangle_J ~ (n \geq 1) \, , \quad J^a_l | j \rangle_J ~ (l \geq 1) \, .
\label{primarycond}
\end{align}
We set the primary states such as to be simultaneously the eigenstate of $L_0$ and in the spin $J$ representation of $J_0^a$  as
\begin{align}
L_0 | j \rangle_J = h | j \rangle_J  \, , \quad J_0^a | j \rangle_J = - (D^{a})^{~i}_{j} | i \rangle_J \, ,
\label{zeromodes}
\end{align}
where $h$ is the conformal weight and $(D^{a})^{~i}_{j}$ is the representation matrix for spin $J$.
We use the convention with $j = 1,2, \ldots , 2 J +1$ and
\begin{align}
J^+_0 |2 J +1 \rangle = 0 \, , \quad J^\pm_n =  J^1_n \pm i J^2_n \, .
\end{align}
The explicit forms of $(D^{a})^{~l}_{j}$  are
\begin{align}
\begin{aligned}
&(D^{1})^{~l}_{j} =  \frac{1}{\sqrt{2}} \left[ \sqrt{j (2 J + 1 - j)}  \delta_{j}^{~ l-1} + \sqrt{l (2 J + 1 - l)} \delta_{j-1}^{~~~l} \right] \, , \\
&(D^{2})^{~l}_{j} =  \frac{1}{\sqrt{2}} \left[ - i \sqrt{j (2 J + 1 - j)}  \delta_{j}^{ ~ l-1} + i \sqrt{l (2 J + 1 - l)} \delta_{j-1}^{~~~l} \right] \, , \\
&(D^{3})^{~l}_{j} = \frac{1}{\sqrt{2}} (2 J - 2 j +2)\delta_{j}^{~ l} \, .
\end{aligned}
\end{align}
In particular, we set $(t^{a})^{~l}_{j} = (D^{a})^{~l}_{j}$ with $J = 1/2$.

An important problem here is how to deal with another set of zero modes, i.e., $Q_0^a$.
They do not commute with $J^a_0$ as in \eqref{CRmodes}, which implies that we cannot use simultaneously the representations of $J_0^a$ and $Q_0^a$.
Since $Q_0^a$ behave as in the spin 1 representation of $su(2)$, the action of $Q_0^a$ to $|j\rangle_J$ yields new states with spin $J-1,J,J+1$. 
For simplicity, here we require that 
\begin{align}
Q_0^a  | j \rangle_J  = w_0 (D^{a})^{~i}_{j} | i \rangle_J 
\label{assumption0}
\end{align}
so that the action of $Q_0^a$ does not yield any other representations.
As we will see below, this assumption works nicely for some restricted cases.

\subsubsection{Null states at level 1}

We would like to find out null states which are both descendant and satisfying the primary conditions \eqref{primarycond}. 
We first consider null states appearing at level 1 with spin $J=0,1/2,1,\ldots$. 
We use the ansatz
\begin{align}
| \chi_{j}^a \rangle_J = ( Q^a_{-1} + e_1 J^a_{-1}  +   i f^{a}_{~bc} e_2 J^b_{-1} J^c_0 + e_3 L_{-1} J^a_0 ) | j \rangle_J \, .
\label{primaryansatz}
\end{align}
If the states are indeed null, then we can consistently set $| \chi_{j}^a \rangle_J  = 0$.
In this case, $Q_{-1}^a$ is given by a linear combination of $J_{-1}^a$ and $L_{-1} J_0^a$.

We first study the case with $J=0$, where the states \eqref{primaryansatz} become simplified as
\begin{align}
| \chi^a_1 \rangle_0 = ( Q^a_{-1} + e_1 J^a_{-1} ) | 1 \rangle_0 \, .
\end{align}
The non-trivial primary conditions are 
\begin{align}
L_1 | \chi^a_j \rangle_J = 0 \, , \quad J_1^a| \chi^b_j \rangle_J = 0  \, , \quad Q^a_1 | \chi^b_j \rangle_J  = 0   \label{level1pc}
\end{align}
with $j=1$ and $J=0$.
The first two conditions  are satisfied with $e_1 = 0$, and then the third condition leads to $h = 0$.
With $h = 0$, the descendant state $L_{-1}| j \rangle_0$ becomes null.
Setting the null states to vanish as $|\chi^a_1 \rangle_0 = Q_{-1} |1 \rangle_0 = 0 $, the primary state $ | 1 \rangle_0$ can be identified with the vacuum satisfying \eqref{vacuumcond}.

The first non-trivial example would be given with $J = 1/2$.
The null states $|\chi^a_j \rangle_{1/2}$ are in the product of the spin $1$ and $1/2$ representations, 
thus we can decompose them into those in the spin $3/2$ and $1/2$ representations. 
For the spin $3/2$ representation, we use
\begin{align}
| \psi_{3/2} \rangle_{3/2} = | \chi^1_{2} \rangle_{1/2} + i | \chi^2_{2} \rangle_{1/2} 
= Q^+_{-1}|2 \rangle_{1/2}  +  (e_1 + e_2) J^+_{-1 } |2 \rangle_{1/2} \, ,
\end{align}
where we have defined
\begin{align}
Q^\pm_n = Q^1_n \pm i Q^2_n \, .
\end{align}
We can equally use those obtained by the action of $(J^-_0)^r$ with $r=1,2$. 
The conditions $ L_1 | \psi_{3/2} \rangle_{3/2} =   J^a_{1}  | \psi_{3/2} \rangle_{3/2} = 0$ set
\begin{align}
e_1 + e_2 = - \frac{w_0}{\ell -1} \, .
\end{align}
Moreover, from $Q^a_1 | \psi_{3/2} \rangle_{3/2} = 0$, we have
\begin{align}
w_0 = \pm i \sqrt{\frac{c_1 (\ell-1) (2 h \ell+8 h +\ell-3)}{2 \ell (4 + 3 \ell)}} \, .
\label{w01}
\end{align}
There is no restriction on $h$ at this stage.
For the  spin $1/2$ representation, we use
\begin{align}
| \psi_{1/2} \rangle_{1/2} = |\chi^1_{1} \rangle_{1/2} + i | \chi^2_{1} \rangle_{1/2} - | \chi^3_{2} \rangle_{1/2} \, .
\end{align}
The conditions $L_1  | \psi_{1/2} \rangle_{1/2}= J^a_{1}  | \psi_{1/2} \rangle_{1/2} = 0$ fix the non-trivial coefficients in terms of $w_0,h$ as
\begin{align}
e_2 - \frac{e_1}{2} =  \frac{(3-4 h) w_0}{4 h (\ell+2)-3} \, , \quad 
e_3 = \frac{4 (\ell+1) w_0}{4 h (\ell+2)-3} \, .
\end{align}
From $Q^a_{1}  | \psi_{1/2} \rangle_{1/2} = 0$, we similarly have
\begin{align}
w_0 = \pm i \sqrt{\frac{c_1 (4 h  (\ell+2)-3) (4 h  (\ell+2) (\ell+4)-\ell (4 \ell+11))}{ 16 \ell \left(3 \ell^2+\ell-4\right) (4 h +3 \ell)}} \, .
\label{w02}
\end{align}
In this way, we find that null states can be constructed from a set of primaries when \eqref{w01} or \eqref{w02} is satisfied.
If both conditions are compatible, then two types of null states are possible from a set of primaries.
Indeed, the two conditions are satisfied for
\begin{align}
h = \frac{7-2 \ell}{4 (\ell+4)}\, ,  \quad h = \frac{-4 \ell^2+6 \ell-5}{4 (\ell-4)} 
\label{h00}
\end{align}
at the same time.
The primary state with the first conformal weight can be realized as a coset state as seen below.

With the experience for $J=1/2$, it is not so difficult to increase $J$.%
\footnote{We have checked the expressions below for $J=1,3/2,\ldots , 3$ but it is not difficult to work with other $J$.}
The states $| \chi^a_{j}\rangle_J$ can be decomposed into those in the representations with spin $J + 1 , J , J - 1$. 
For  spin $J+1$, we use
\begin{align}
| \psi_{J+1} \rangle_{J+1} = | \chi^1_{2J+1} \rangle_{J} + i | \chi^2_{2J+1} \rangle_{J} \, .
\end{align}
The conditions \eqref{level1pc} are satisfied by
\begin{align}
e_2 + \frac{e_1}{2 J} = - \frac{w_0}{\ell - 2 J} \, , \quad
w_0 = \pm  i \frac{ \ell-2 J}{2 J} \sqrt{\frac{c_1(h (\ell+4)+J (-2 J+\ell-2))}{\ell \left(3 \ell^2+\ell-4\right)}} \, .
\label{w11}
\end{align}
For spin $J$, we similarly use
\begin{align}
| \psi_{J} \rangle_{J} = | \chi^3_{2J+1} \rangle_{J} - \frac{1}{\sqrt{2J}}  ( | \chi^1_{2J}  \rangle_{J}  + i | \chi^2_{2J} \rangle_{J} ) \, ,
\end{align}
and the primary conditions \eqref{level1pc} lead to
\begin{align}
\label{w12}
&e_2 - \frac{e_1}{2} = \frac{w_0 \left(- h+J^2+J\right)}{h (\ell+2)-J (J+1)} \, , \quad
e_3 = \frac{(\ell+1) w_0}{h (\ell+2)-J (J+1)} \, , \\
& w_0 = \pm i \frac{  \sqrt{c_1 \left( h^2 (\ell+4) (\ell+2)^2- h (\ell+2)^2 (2 J (J+1)+\ell)+J (J+1) \ell \left(J^2+J+\ell+2\right) \right)}}{2 \sqrt{\ell \left(3 \ell^2+\ell-4\right) (h +J (J+1) \ell)}} \, . \nonumber
\end{align}
For spin $J-1$, we set
\begin{align}
\begin{aligned}
| \psi_{J-1} \rangle_{J-1} = &| \chi^1_{2J+1} \rangle_{J} - i | \chi^2_{2J+1} \rangle_{J} + \frac{2}{\sqrt{2 J}}| \chi^3_{2J} \rangle_{J}  \\
&  - \frac{1}{\sqrt{J( 2J - 1)}}  ( | \chi^1_{2J-1}  \rangle_{J}  + i | \chi^2_{2J-1} \rangle_{J} ) \, ,
\end{aligned}
\end{align}
then we  find 
\begin{align}
\begin{aligned}
&e_2  - \frac{e_1}{2 J +2} = - \frac{w_0}{\ell + 2 J + 1} \, , \\
&w_0 = \pm i \frac {(2 J+\ell+2)}{(2 J+2)} \frac{ \sqrt{ c_1 ( 2 h (\ell+4)-(2 J+2) (2 J+\ell))}}{ \sqrt{2 \ell \left(3 \ell^2+\ell-4\right)}}
\end{aligned}
\label{w13}
\end{align}
as solutions to the primary conditions \eqref{level1pc}.

In this way, we have three types of null states for generic $J$.
However, in general, there is no $h$ which satisfies all of \eqref{w11}, \eqref{w12} and \eqref{w13} simultaneously.
Therefore, we have to give up at least one of them.
The maximal number of null state can be obtained if both \eqref{w11} and \eqref{w12} are satisfied. 
This condition  leads to
\begin{align}
h = \frac{J (3 J-\ell+2)}{\ell+4} \, ,  \quad  h =  \frac{J \left(3 J^2-3 J \ell+J+\ell^2\right)}{J (\ell+4)-\ell} \, .
\label{nullcw}
\end{align}
Setting $J =1/2$, this reproduces \eqref{h00}.
Furthermore, we will see that the state with the first conformal weight can be realized as a coset state in the spin $J$ representation of $su(2)$.

We may relax the condition for the maximal number of null states.
Instead of \eqref{w11} and \eqref{w12}, we may require that  \eqref{w11} and \eqref{w13} are satisfied.
Then we find
\begin{align}
h = \frac{J (J+1) (12 J (J+1)+(\ell-2) \ell)}{(\ell+4) (4 J (J+1)-\ell)} \, . \label{nullcw2}
\end{align}
Similarly, the choice of  \eqref{w12} and \eqref{w13} leads to
\begin{align}
h =  \frac{(J+1) (3 J+\ell+1)}{\ell+4} \, ,  \quad  h =  \frac{(J+1) \left(3 J^2+3 (J+1) \ell+5 J+\ell^2+2\right)}{J (\ell+4)+2 (\ell+2)} \, . \label{nullcw3}
\end{align}
Currently, we do not have any good interpretations of them.

\subsubsection{Null states at level 2}

We also consider null states at level 2 for simple examples with $J=0,1/2$.
For this, we use the ansatz
\begin{align}
\begin{aligned}
| \chi_j \rangle_J = (L_{-2} + g_1 L_{-1} L_{-1} + g_2 Q^a_{-2} J^a_{0} + g_3 J^a_{-2} J^a_0 + g_4 Q^a_{-1} L_{-1} J^a_0 + g_5 L_{-1} J^a_{-1} J^a_0 \\ + g_6 Q^a_{-1} Q^a_{-1} + g_7 Q^a_{-1} J^a_{-1} + g_8 J^a_{-1} J^a_{-1} + i f_{abc} g_9  Q^a_{-1} J^b_{-1} J^c_{0} )| j \rangle_J \, ,
\label{nulllevel2}
\end{aligned}
\end{align}
where we have set the state of the form $L_{-2}|j\rangle_J + \cdots$.
The states become null if the non-trivial primary conditions
\begin{align}
L_r | \chi^a_j \rangle_J= 0 \, , \quad J_r^a| \chi^b_j \rangle_J = 0  \, , \quad Q^a_r | \chi^b_j \rangle_J  = 0   \label{level2pc}
\end{align}
with $r=1,2$ are satisfied.
Since the requirement \eqref{assumption0} leads to $(G_0^a + w_0 J_0^a )|j \rangle_J = 0$,
the states
\begin{align}
G^a_{-2} |j \rangle_J  + \cdots =  \frac{1}{2} [ G^a_0 + w_0 J_0^a  ,L_{-2}  ] |j \rangle_J  + \cdots 
\end{align}
become null if the states \eqref{nulllevel2} satisfy the primary conditions \eqref{level2pc}.

For $J=0$, the ansatz becomes simplified as
\begin{align}
\begin{aligned}
| \chi_a \rangle_J = (L_{-2} + g_1 L_{-1} L_{-1}  + g_6 Q^a_{-1} Q^a_{-1} + g_7 Q^a_{-1} J^a_{-1} + g_8 J^a_{-1} J^a_{-1}  )| j \rangle_J \, .
\end{aligned}
\end{align}
The primary conditions \eqref{level2pc} determine the coefficients $g_1,g_6,g_7 ,g_8$ in terms of $h,\ell$.
There are three solutions for the conformal weight $h$ as
\begin{align}
h =  \frac{1-\ell}{\ell+4} \, , \quad h =  \frac{\ell+1}{\ell+4} \, , \quad h =  \frac{1}{4} (3 \ell+4) \, . \label{nullcw4}
\end{align}
The state with the last conformal weight behaves as $\mathcal{O}(\ell)$ for large $\ell$, and it will be related to a conical defect geometry of \cite{Castro:2011iw} below. 

For $J=1/2$, the primary conditions \eqref{level2pc} fix the coefficients $g_i$ $(i= 1,2,\ldots,9)$ in terms of $h ,w_0 , \ell$.
Moreover, $h$ and $w_0$ are also determined as
\begin{align}
h = \frac{39}{4 (\ell+4)}-\frac{3}{2} \, , \quad w_0 = - i \frac{  \sqrt{c_1 (\ell-3)^2}}{2 \sqrt{\ell \left(3 \ell^2+\ell-4\right)}} \, . \label{nullcw5}
\end{align}

\subsection{Hamiltonian reduction of $sl(4)$}
\label{sec:hamilton}

The degenerate representations of the  W-algebra may be examined from those of the $sl(4)$ Wess-Zumino-Novikov-Witten (WZNW) model by applying the Hamiltonian reduction.
As argued in \cite{Creutzig:2018pts}, we may study the Hamiltonian reduction  by following the procedure of  \cite{Creutzig:2015hla} (see also \cite{Hikida:2007tq,Hikida:2007sz,Creutzig:2011qm}).
It is convenient to use the description of the $sl(4)$ WZNW model as (see (4.6) of \cite{Creutzig:2015hla})
\begin{align}
\begin{aligned}
S_t [\phi , g_1 , g_2 , \gamma , \bar \gamma , \beta , \bar \beta] 
=& S^\text{WZNW}_{t+2} [g_1] + S^\text{WZNW}_{t+2} [g_2] + \frac{1}{2 \pi } \int d^2 z \left[ \partial \phi \bar \partial \phi - b  \sqrt{g}  \mathcal{R} \phi \right] 
\\ &+ \frac{1}{2 \pi} \int d^2 z\,  \text{tr} (\beta \bar \partial \gamma + \bar \beta \partial \bar \gamma - \frac{1}{t} e^{- 2 \phi} \bar \beta g_1^{-1} \beta g_2 ) \, .
\end{aligned}
\end{align}
Here $\gamma , \bar \gamma , \beta , \bar \beta$ are $2 \times 2$ matrix valued fields and 
$b = - i /\sqrt{2 (t + 4)}$. We consider the vertex operators of the form
\begin{align}
V_{\alpha ,\lambda_1 ,\lambda_2} = e^{2 \alpha b \phi} V_{\lambda_1}^{sl(2)} (g_1 )V_{\lambda_2}^{sl(2)} (g_2 ) \, , 
\label{vo}
\end{align}
where $\lambda_i$ label the representations of $sl(2)$. The conformal weight of this operator is
\begin{align}
h = \frac{C_2 (\lambda_1)}{t+4} + \frac{C_2 (\lambda_2)}{t+4} - b^2 \alpha (  \alpha + 4  ) \, .
\label{cw1}
\end{align}
Here $C_2(\lambda_i)$ is the second Casimir of $sl(2)$ for the representation $\lambda_i$.
After the reduction procedure, the action becomes (see (3.5) of \cite{Creutzig:2018pts})
\begin{align}
\begin{aligned}
S_t [\varphi ,g_1 ,g_2] =& S^\text{WZNW}_{t + 2} [g_1] + S^\text{WZNW}_{t + 2} [g_1]  \\
&+ \frac{1}{2 \pi} \int d^2 z \left[ \partial \varphi \bar \partial \varphi + \frac{Q_{\varphi}}{4} \sqrt{g} \mathcal{R} \varphi + \frac{1}{t} \text{tr} \left(e^{- 2 b \varphi} g_1^{-1} g_2 \right)  \right ]
\end{aligned}
\end{align}
with
\begin{align}
Q_\varphi = - 4 b - 1/b \, .
\end{align}
The conformal dimension of the vertex operator \eqref{vo} is
\begin{align}
h = \frac{C_2 (\lambda_1)}{t+4} + \frac{C_2 (\lambda_2)}{t+4} - b^2 \alpha ( \alpha + 4 ) - \alpha \, .
\label{cw2}
\end{align}
Thus the shift of conformal dimension is $\delta h = - \alpha$.

Next we examine the cases with finite dimensional representations of $sl(4)$.
We express the representation of $sl(4)$ by a Young diagram $\Lambda '$ with three integers $(\lambda_1 ' , \lambda_2 ' , \lambda_3 ')$. We also define $l_j ' = \sum_{s=j}^3 \lambda_{s} ' $, where $l_j '$ counts the number of boxes of $\Lambda '$.
With the orthogonal basis $\epsilon_j$ $(j=1,2,3,4)$, the highest weight for the representation is
\begin{align}
\Lambda ' = \sum_{j =1}^3 l '_j \epsilon_j  -\frac{|\Lambda '|}{4} \sum_{j=1}^4 \epsilon_j  
\end{align}
with $|\Lambda ' | = \sum_{j=1}^3 l '_j$. 
The second Casimir of $sl(4)$ for the representation $\Lambda '$ is
\begin{align}
C_2 (\Lambda ') = \frac12  \sum_{j=1}^3 ( l '_j )^2 - \frac{|\Lambda '|^2 }{4} - \sum_{j=1}^4 j l ' _j + \frac{5 |\Lambda '| }{2} \, . \label{Casimirsl4}
\end{align}
In order to use the vertex operator in \eqref{vo}, we need to rewrite the label  in terms of $sl(2) \oplus sl(2) \oplus u(1)$. We may express the highest weights for the representations as
\begin{align}
\Lambda^{(1)} = \frac{\lambda_1}{2} \epsilon_1 - \frac{\lambda_2}{2} \epsilon_2 \, , \quad
\Lambda^{(2)} = \frac{\lambda_2}{2} \epsilon_3 - \frac{\lambda_2}{2} \epsilon_4 \, , \quad
\bar m = \frac{m}{8} (\epsilon_1 + \epsilon_2 - \epsilon_3 - \epsilon_4) \, .
\end{align}
The relation to the highest weight for $sl(4)$ is 
\begin{align}
\lambda_1 ' = \lambda_1  \, , \quad \lambda_2 ' = - \frac{\lambda_1}{2} - \frac{\lambda_2}{2} + \frac{m}{4}  \, , \quad 
\lambda_3 ' = \lambda_2 \, .
\label{l13m}
\end{align}
Using \eqref{Casimirsl4}, the conformal weight for the $sl(4)$ WZNW model is computed as
\begin{align}
h = \frac{C_2 (\Lambda ')}{t + 4} = \frac{ 8 \lambda_1 ( \lambda_1 + 2) + 8 \lambda_2 ( \lambda_2 + 2) + m (m + 16) }{32 (t + 4)} \, .
\end{align}
Compared with \eqref{cw1}, we find $\alpha = m/4$.
Thus, the conformal weight for the W-algebra is obtained as
\begin{align}
h = \frac{8 \lambda_1 ( \lambda_1 + 2) + 8 \lambda_2 ( \lambda_2  + 2) + m ( m + 16)  }{32 ( t + 4) } - \frac{m}{4} 
\label{cw3}
\end{align}
from \eqref{cw2}.
We would like to claim that it is the conformal weight of the primary operator belonging to a degenerate representation of the W-algebra.

Let us examine several examples and compare them with previous results.
For this, it might be convenient to use the level $\ell$ of $sl(2)$ currents instead of the level $t$ of $sl(4)$.
The relation is $t = \ell/2 - 2$, see \eqref{DSl}.
We first set
\begin{align}
(\lambda_1 , \lambda_2 , m) = (2 J , 0 , 4 J ) \, ,
\end{align}
then we find
\begin{align}
h = \frac{J (3 J - \ell + 2)}{\ell + 4} \, .
\label{DScw}
\end{align}
This is the first conformal weight in \eqref{nullcw}.
Next we set
\begin{align}
(\lambda_1 , \lambda_2 , m) = (2 J , 0 , 4 J  + 4) \, ,
\label{l12mex}
\end{align}
then the conformal weight becomes
\begin{align}
h = \frac{(J +1)(3 J - \ell + 1)}{\ell + 4} \, .
\end{align}
This is given by the first equation of \eqref{nullcw4} for $J=0$ and \eqref{nullcw5} for $J =1/2$.

We may consider the case with $t +4 = 2 p/q$, where $p,q$ are co-prime with each other.
According to \cite{Kac:1988qc}, the characters of the highest weight with
\begin{align}
\lambda_i ' = (1 - r_i ') (t +4) - (1 - s_i ') \, , \quad 1 \leq s_i ' \leq 2 p -1 \, , \quad 1 \leq r_i ' \leq q
\end{align}
transform with each other under the modular transformation.
Thus we may put
\begin{align}
\begin{aligned}
&\lambda_1 = (1 - r_1 ' ) (t +4) - (1 - s_1 ' ) \, , \\
&\lambda_2 = (1 - r_3 ') (t +4) - (1 - s_3 ') \, , \\
& m = 2 (4 - r_1 ' - 2 r_2 ' - r_3 ') (t +4) - 2 (4 - s_1  ' - 2 s_2 ' - s_3 ') 
\end{aligned}
\end{align}
into the expression of \eqref{cw3}.
For $\lambda_1 = \lambda_2 = 0$ and $m  = - 4 (t + 4)$,
we find
\begin{align}
h = \frac{1}{4} (3 \ell + 4) \, , 
\end{align}
which reproduce the last equation of \eqref{nullcw4}.
More generically, we may set
\begin{align}
\lambda_1 = - n_1 (t + 4) \, , \quad \lambda_2 = - n_2 (t +4) \, , \quad  m = 4(1 - n_3 ) (t + 4)  
\end{align}
with
\begin{align}
n_1 , n_2 = 0 ,1,2,\ldots , q -1 \, , \quad  n_3 - (n_1 + n_2)/2 = 1 , 2 ,\ldots , q \, . \label{nicond}
\end{align}
Then we find
\begin{align}
h = \frac{ l}{8} \left(n_1^2+n_2^2+2 n_3^2-2\right)+\frac{1}{2} \left((n_1-1) n_1 +(n_2-1) n_2+2 (n_3-1)^2\right) \, . \label{cw4}
\end{align}
These conformal weights are of order $\mathcal{O}(\ell)$ and the corresponding states may be mapped to 
conical defect solutions of \cite{Castro:2011iw} 
for the $sl(4) \oplus sl(4)$ Chern-Simons gravity as seen in subsection \ref{sec:conical}.

\subsection{Spectrum of the dual coset}
\label{sec:cosetspe}

We have argued that the rectangular W-algebra can be realized as the symmetry algebra of the coset \eqref{coset}  in subsection \ref{sec:dualcoset}.
In this subsection, we compare the results obtained in the previous subsections to the coset states.
The primary states in the coset \eqref{coset} are labeled by $(\Lambda_{N+M}; \Lambda_N , m)$, where $\Lambda_L$ represents the highest weight of $su(L)$ and $m \in \mathbb{Z}_\kappa$ is the $u(1)$ charge.
The conformal weight of the state $(\Lambda_{N+M}; \Lambda_N , m)$ is
\begin{align}
h = n + h^{N+M,k}_{\Lambda_{N+M}} - h_{\Lambda_N}^{N,k} - h^\kappa_m \, , \label{cwformula}
\end{align}
where $n$ is an integer related to how the representations for the denominator are embedded in those for the numerator, see, e.g., \cite{DiFrancesco:1997nk}. We express the conformal dimensions of primaries for $su(L)_K$ and $u(1)_\kappa$ as
\begin{align}
h_{\Lambda_L}^{L,K} = \frac{C_2 (\Lambda_L)}{K +L} \, , \quad
h^\kappa_m = \frac{m^2}{2 \kappa} \, , \label{cwformulasub}
\end{align}
respectively.
Here $C_2 (\Lambda_L)$ is the second Casimir of $su(L)$ for the representation $\Lambda_L$.

As discussed in \cite{Creutzig:2013tja,Hikida:2015nfa}, basic states may be 
\begin{align}
(\text{f}; 0 , N) \otimes (\bar{\text{f}}; 0 ,  - N) \, , \quad 
(0; {\text{f}} ,- N - M) \otimes (0;\bar {\text{f}} , N + M) \, ,
\end{align}
where the holomorphic and anti-holomorphic parts are combined in the charge conjugated manner.
Here we express  the fundamental and anti-fundamental representations by $\text{f}$ and $\bar{\text{f}}$, respectively.
The conformal weights are given by
\begin{align}
h =\frac{k (M (M+N)-1)-N}{2 k M (k+M+N)} \, , \quad 
h =-\frac{-2 k^2 M-k M N+k+M+N}{2 k^2 M+2 k M N} \, .
\end{align}
Using $\lambda = k/(k +N)$ as in \eqref{tHooft}, we find
\begin{align}
h = \frac{M (- k \lambda +k+\lambda  M)-1}{2 M (k+ \lambda M)} \, , \quad 
h =  \frac{1}{2} \left(-\frac{\lambda }{k^2}-\frac{1}{k M}+ \lambda +1\right) \, .
\label{h12tHooft}
\end{align}
These two expressions are exchanged if we use an alternative definition of $\lambda$.
These states were proposed to be dual to two complex scalars $\phi^{j \bar \jmath}_i$ $(i=1,2)$, which transform as in the (anti-)fundamental representation of $su (2)$ in the (anti-)holomorphic sector. A set of states can be generated by fusing these basic ones, and they are supposed to be dual to composite bulk fields.

In order to compare with the previous results, we set $\lambda =2$ and $M=2$.
We can see that the first expression of \eqref{h12tHooft} reduces to \eqref{h00}. 
Generically, we should consider a state in the spin $J$ representation of $su(2)$. 
Among those constructed by fusing the basic states, we consider the states labeled by
\begin{align}
((2 J , 0 , \ldots ,0) ; 0 , 2 J N) \, ,
\end{align}
where $\Lambda_{N+2} =(2 J , 0 , \ldots ,0) $ denotes the $2 J$-th symmetric representation of $su(N+2)$. 
From  formula \eqref{cwformula} with \eqref{cwformulasub}, we can compute the conformal weight of the state as
\begin{align}
h = \frac{J (3 J-k+2)}{k+4} \, .
\end{align}
This precisely reproduces the first conformal weight in \eqref{nullcw} or \eqref{DScw}.

\subsection{Conical defect geometry}
\label{sec:conical}

In the above analysis, we have found some states with conformal weight of order $\ell$.
The parameter $\ell$ becomes large when the dual gravity is in the classical regime.
As in \cite{Castro:2011iw,Gaberdiel:2012ku,Perlmutter:2012ds,Hikida:2012eu}, we would like to interpret the states with conformal weight \eqref{cw4} in terms of classical geometry of $sl(4) \oplus sl(4)$ Chern-Simons gauge theory. Conical defect geometry was discussed even in the cases with non-principal embedding of $sl(2)$ \cite{Castro:2011iw}.
We consider classical solutions expressed as
\begin{align}
A = e^{- \rho L_0} a  e^{\rho L_0} dz  + d \rho \, , \quad \bar A = e^{\rho L_0} \bar a e^{- \rho L_0} d \bar z - d \rho \, .
\end{align}
Here $(z , \bar z , \rho)$ are the bulk coordinates, $a, \bar a$ are $sl(4)$-valued constants, and $L_0\, (\equiv x)$ is a generator of the embedded $sl(2)$.

Non-singular solutions were proposed to be characterized by the trivial condition of the holonomy matrix $\exp ( \oint d \phi \,  a )$ with $z = \phi + i t$ $(\phi \sim \phi + 2 \pi)$.
After the diagonalization, we may express $a$ as
\begin{align}
a = i n_1 
\begin{pmatrix} 
\frac12 & 0 & 0 & 0 \\ 
0 & - \frac12 & 0 & 0 \\
0 & 0 & 0 & 0 \\
0 & 0 & 0 & 0
\end{pmatrix}
+  i n_2 
\begin{pmatrix} 
0 & 0 & 0 & 0 \\ 
0 & 0 & 0 & 0 \\
0 & 0 & \frac12 & 0 \\
0 & 0 & 0 & - \frac12 
\end{pmatrix}
+ i n_3 
\begin{pmatrix} 
\frac12 & 0 & 0 & 0 \\ 
0 &  \frac12 & 0 & 0 \\
0 & 0 & - \frac12 & 0 \\
0 & 0 & 0 & - \frac12 
\end{pmatrix} \, .
\end{align}
Here, the matrices correspond to the generators $t^3 \otimes (\mathbbm{1}/2 + t^3)$, $t^3 \otimes (\mathbbm{1}/2 - t^3)$, and $\mathbbm{1} \otimes t^3$.
We require that  the holonomy matrix $\exp ( \oint d \phi \,  a )$ becomes a center of $SU(4)$, that is $\mathbb{Z}_4 $. This leads to the condition $n_1 , n_2 \in \mathbb{Z}$ and $n_3 + (n_1 + n_2) /2 \in \mathbb{Z}$. Following \cite{Castro:2011iw}, we assign the condition such that the matrix $a$ is not degenerate, and this leads to $n_3 \neq 0$ and $- n_1 + n_3 \neq n_2 - n_3$. 
Using the symmetry of $SU(4)$, we can set $n_1 \geq 0 , n_2 \geq 0$, and moreover choose $n_i$ to satisfy $- n_1 + n_3 > n_2 - n_3$. With this choice, the parameters $n_i$ now take
\begin{align}
n_1 , n_2 = 0 ,1,2,\ldots  \, , \quad  n_3 - (n_1 + n_2)/2 = 1 , 2 ,\ldots \, . \label{nicond2}
\end{align}
This reproduces \eqref{nicond} except for the upper bounds, which are usually not visible from classical geometry analysis.

The dual conformal weight can be computed as
\begin{align}
h = \frac{c}{24 \epsilon_P} \text{tr} (a^2 ) + \frac{c}{24} \, .
\end{align}
The central charge $c$ and the normalization constant $\epsilon_P$ are (see \cite{Castro:2011iw,Creutzig:2018pts})
\begin{align}
c = 12 k_\text{CS} \epsilon_P = - 6 \ell \, , \quad \epsilon_ P = \frac{1}{12} M n (n^2 -1) = 1 \, . 
\end{align}
Here we considered $sl(Mn)$ with $M=n=2$ and used $k_\text{CS}$ as the level of the Chern-Simons theory. 
The level of $su(2)$ currents is $\ell = - n k_\text{CS} = - 2 k_\text{CS}$.
In summary, we have
\begin{align}
h = \frac{\ell}{8} \left(n_1^2 + n_2 ^2 + 2 n_3 ^2  -2  \right) \, , 
\end{align}
where $n_i$ take values in \eqref{nicond2}.
This expression reproduces \eqref{cw4} at the leading order in $1/\ell$.

\section{$\mathcal{N}=2$ rectangular W-algebra}
\label{sec:N=2}

In previous sections, we have analyzed the rectangular W-algebra, which appears as the asymptotic symmetry of higher spin gravity with $M \times M$ matrix valued fields. In order to relate higher spin gravity to superstring theory, it is important to introduce supersymmetry. In \cite{Creutzig:2011fe}, a holography involving the $\mathcal{N}=2$ higher spin supergravity of \cite{Prokushkin:1998bq} was proposed without the matrix extension. The classical asymptotic symmetry of the higher spin supergravity has been analyzed in \cite{Creutzig:2011fe,Hanaki:2012yf,Henneaux:2012ny}.
The gauge algebra of the higher spin supergravity is given by $shs[\lambda]$, which can be truncated to $sl(n+1|n)$ at $\lambda = - n$, see, e.g., \cite{Bergshoeff:1991dz} for some details of $shs[\lambda]$.
The gauge algebra for the matrix extension is denoted as $shs_M[\lambda]$ \cite{Gaberdiel:2013vva,Creutzig:2013tja}, which can be reduced to $sl(M(n+1)|Mn)$ at $\lambda = - n$.
The holography with the extended higher spin supergravity was proposed also in \cite{Creutzig:2013tja}.

The Lie superalgebra $sl(M (n+1)| M n)$ can be decomposed as 
\begin{align}
sl(M (n+1)| M n) \simeq sl(M) \otimes \mathbbm{1}_{n+1|n} \oplus \mathbbm{1}_M \otimes sl(n+1|n) \oplus sl(M) \otimes sl(n+1 | n)  \label{decssl}\, .
\end{align}
The generators of $sl(R|S)$ can be expressed by $(R +S) \times (R +S)$ supermatrices, and $\mathbbm{1}_{R|S}$ denotes
\begin{align}
\mathbbm{1}_{R|S} = 
\begin{pmatrix}
\mathbbm{1}_R & 0 \\
0 & 0 
\end{pmatrix}
\oplus
\begin{pmatrix}
0 & 0 \\
0 & \mathbbm{1}_S
\end{pmatrix} \, .
\end{align}
Three dimensional $\mathcal{N}=1$ supergravity on AdS can be described by $osp(1|2)$ Chern-Simons gravity \cite{Achucarro:1987vz}.  We use  the $osp(1|2)$ principally embedded  in $\mathbbm{1}_M \otimes sl(n+1|n)$ as the supergravity sector.
With the action of $sl(2) \subset osp(1|2)$, the gauge algebra can be decomposed as
\begin{align}
\begin{aligned}
sl(M(n+1)|M n) \simeq \, &  sl(M) \oplus  sl(M) \oplus u(1)\\
& \oplus   2 M^2  \left( \bigoplus_{s=2}^{n+1} g^{(s-1/2)} \right)   \oplus 2 M^2  \left( \bigoplus_{s=2}^{n} g^{(s)}   \right) \oplus M^2  g^{(n+1)} \, .
\end{aligned}
\end{align}
After the Hamiltonian reduction, the spectrum consists of two $sl(M)$ currents, one $u(1)$ current, $2M^2$ spin $s$ currents with $s=2,3,\ldots ,n$ and $M^2$ spin $n+1$ currents. There are also $2 M^2$ fermionic currents with spin $s-1/2$ with $s=2,3,\ldots,n+1$. One of the spin 2 currents is given by the energy-momentum tensor.

In the next subsection, we compute the central charge $c$ of the algebra and the levels $\ell_1,\ell_2$ for the two sets of $su(M)$ currents by applying the method in subsection \ref{sec:basic}.
In subsection \ref{sec:sOPE}, we compute the OPEs among generators by requiring their associativity.
We work on the simple case with $n=1$, where the algebra includes only spin $1,3/2,2$ generators.
In subsection \ref{sec:sminimal}, we examine the degenerate representations with level 1 null vectors and primary states in the spin $1/2$ representation of $su(2)$. For this, we apply the method developed in appendix \ref{sec:minimal}.
In subsection \ref{sec:sdualcoset}, we compare the results obtained so far to those of dual coset \eqref{scoset}.

\subsection{Central charge and levels of the affine symmetries}
\label{sec:sbasic}

We start by computing the central charge $c$ of the algebra by applying the formula of \eqref{KW}.
The superdimension of $sl(M(n+1)|M n)$ is 
\begin{align}
\begin{aligned}
\text{sdim} \, sl(M(n+1)|M n) &= M^2 (n +1)^2 -1 + M^2 n^2 - 1 + 1 - 2 M^2 n (n+1) \\
&= M^2 -1 \, ,
\end{aligned}
\end{align}
and the dual Coxeter number is $h^\vee = M (n+1) - M n = M$. The norm of $x$ for the $sl(2) \subset osp(1|2)$ is
\begin{align}
(x|x) = M\cdot \left[ \frac{1}{12} (n+1)((n+1)^2 -1) - \frac{1}{12} n (n^2 -1) \right] = \frac{1}{4}M n (n + 1)  \, .
\end{align}
There are $M^2 ((n +1 - j) + (n - j))$ sets of fermionic ghost system for $j=1,2,\ldots ,n$ and $ 2 M^2 (n - l)$ sets of bosonic ghost system for $j = l + 1/2$ with $ l = 0 , 1 , \ldots , n-1$. 
Using formula \eqref{KW}, the total central charge is obtained as
\begin{align}
c = \frac{t (M^2 -1)}{t + M} - 3 t M n (n+1) - 3 M^2 n^2 \, . \label{sDSc}
\end{align}

We then compute the levels $\ell_1,\ell_2$ for the two $su(M)$ currents.
A set of $su(M)$ currents comes from $sl(M) \otimes \mathbbm{1}_{n+1} $, where $\mathbbm{1}_{n+1}$ corresponds to the identity in $sl(n+1) \subset sl(n+1| n) \subset sl(M(n+1)|Mn)$.
Therefore, the ghosts from  $sl(M(n+1)) \subset sl(M(n+1)|Mn)$ give rise to the shift of level by $Mn (n+1) $.
Similarly, the other set comes from $sl(M) \otimes \mathbbm{1}_{n}$, and the ghosts from $sl(M n ) \subset sl(M(n+1)|Mn)$ contribute to the shift of level by $M n (n-1)$.
There are bosonic ghosts arising from the off-diagonal blocks of $sl(M(n+1)|Mn)$, 
which transform in the bi-fundamental representation of $sl(n+1) \oplus sl(n)$ and in the trivial and adjoint representation of $sl(M)$.
A set of bosonic ghosts in the adjoint representation of $sl(M)$ yields the shift of level by $-M$.
Thus the shifts of level are $-M n ( n+1)$ for both $sl(M)$.
There are also $n$ fermionic ghost systems of conformal weight $(1/2,1/2)$ in the trivial and adjoint representations of $sl(M)$, and the contributions to the levels are $M n$. 
In total, the levels of two sets of $sl(M)$ currents are
\begin{align}
\begin{aligned}
\ell_1& =   t (n +1) +  M n (n+1) - M n ( n+1) + M n =  t (n + 1) + M n   \, , \\
\ell_2 &= - t n + M n (n -1)- M n ( n+1) + M n = - t n - M n \, .
\end{aligned} \label{sDSl}
\end{align}
Since $t$ plays no role in the $\mathcal{N}=2$ W-algebra, we may remove it using the above expression of $\ell_1$. The central charge $c$ and the other level $\ell_2$ can be written as
\begin{align}
c = \frac{\ell_1 \left(M^2-1\right)-M n \left(3 \ell_1^2+3 \ell_1 M+M^2-1\right)}{\ell_1+M} \, , \quad \ell_2 = -\frac{n (\ell_1+M)}{n+1} \label{cell2}
\end{align}
in terms of $\ell_1$.

\subsection{OPEs among generators}
\label{sec:sOPE}

As explained above, the $\mathcal{N}=2$ rectangular W-algebra is generated by a $u(1)$ current $K$, two sets of  $su(M)$ currents $J^a,K^a$  $(a=1,\ldots, M^2 -1)$ as well as  bosonic and fermionic higher spin currents.
The $u(1)$ current satisfies
\begin{align}
K (z) K(0) \sim \frac{\kappa_K}{3 z^2} \, , 
\label{KKOPE}
\end{align}
where the normalization constant $\kappa_K$ will be fixed later.
The OPEs of $su(M)$ currents are
\begin{align}
J^a (z) J^b (0) \sim \frac{\ell_1 \delta^{ab}}{z^2} + \frac{i f^{ab}_{~~c} J^c (0)}{z} \, , \quad 
K^a (z) K^b (0) \sim \frac{\ell_2 \delta^{ab}}{z^2} + \frac{i f^{ab}_{~~c} K^c (0)}{z} 
\label{JJKKOPE}
\end{align}
with levels $\ell_1,\ell_2$.
The algebra includes the energy-momentum tensor satisfying
\begin{align}
T(z) T(0) \sim \frac{c/2}{z^4} + \frac{2 T(0)}{z^2} + \frac{ T ' (0)}{z} 
\end{align}
with central charge $c$. For a moment, we do not specify the levels $\ell_1,\ell_2$ and the central charge $c$.
We denote the bosonic currents of spin $s$ by $W^{-(s),A}$ $(s=1,2,\ldots ,n)$ and $W^{+(s),A}$ $(s=2,3,\ldots,n+1)$ with $A = ( 0 ,a )$ along with $J^a$. Here we set
\begin{align}
W^{-(1),0} \equiv K \, , \quad W^{-(1),a} \equiv K^a \, , \quad W^{+(2),0} \equiv T \, .
\end{align}
The fermionic currents of spin $s$ are represented by $G^{\pm(s),A}$ $(s=3/2,5/2,\ldots,n+1/2)$.
We use the basis for these fields primary w.r.t.~the Virasoro algebra; that is,
\begin{align}
&T (z) W^{\pm (s),A} (0) \sim \frac{s W^{\pm (s), A}(0)}{z^2} +  \frac{{ W^{\pm (s), A}} ' (0)}{z} \, , \quad
T (z) J^{a} (0) \sim \frac{ J^a (0)}{z^2} +  \frac{{ J^a} ' (0)}{z} \, , \nonumber \\
&T(z) G^{\pm (s) , A} (0) \sim \frac{s G^{\pm (s), A}(0)}{z^2} +  \frac{ { G^{\pm (s), A} } ' (0)}{z} \, .
\end{align}

\subsubsection{Composite primary operators}

We determine the OPEs among generators by requiring the associativity.
We only consider the case with $n=1$, where the spin content of the algebra is $s=1,3/2,2$.
Thus, the operator products produce (composite) operators only up to spin 3, and we list all of the possible composite operators.

We start by counting the number of independent primary operators.
We do this by decomposing the vacuum character of the $\mathcal{N}=2$ W-algebra 
in terms of the Virasoro characters, see \cite{Candu:2012tr} for the case without matrix extension.
The vacuum character of the $\mathcal{N}=2$ W-algebra is
\begin{align}
\chi^{\mathcal{N}=2,M}_\infty (q) = \prod_{s=1}^n \prod_{i=s}^\infty \left[ \frac{( 1 + q^{i + 1/2})^2 }{(1 - q^i) (1 - q^{i+1})}\right]^{M^2} \prod_{i=1}^\infty \left[\frac{1 }{1 - q^i} \right]^{M^2 -1} \, ,
\end{align}
and the decomposition is
\begin{align}
\chi^{\mathcal{N}=2,M}_\infty (q) = \chi_0 (q) + \sum_{i= 2}^\infty d(i/2) \chi_{i/2} (q) \, .
\label{dec}
\end{align}
Here the Virasoro characters can be found in \eqref{Virch}, and $d(i/2)$ counts the number of independent primary operators with conformal weight $i/2$.
Expanding \eqref{dec} in $q$, we find
\begin{align}
\begin{aligned}
&d(1) = 2 M^2 -1 \, , \quad d(3/2) =2 M^2  \, , \quad d(2) = 2 M^4 -1 \, , \\
&d(5/2) = 2 (2 M^4 - M^2) \, , \quad d(3) = \tfrac{2}{3} (2 M^6 + 9 M^4 - 11 M^2 + 3)  
\end{aligned}
\label{dh}
\end{align}
with $n=1$.

We explicitly construct composite operators primary w.r.t.~the Virasoro algebra. 
With $n=1$, the algebra is generated by spin 1 currents $J^a,K^a,K$, spin 3/2 currents $G^\pm \equiv G^{\pm (3/2),0}$, $G^{\pm,a} \equiv G^{\pm (3/2),a}$, and spin 2 currents $T,Q^a \equiv W^{+ (2),a}$.
We use the abbreviated notation of composite operators as in \eqref{ab}.
We can see that there are no composite primary operators for spin 1 and 3/2 currents.
We find
\begin{align}
[J^{(a} J^{b)}] \, , \quad [K^{(a} K^{b)}]\, , \quad [J^{a} K^b]  \, , \quad [K J^a] \, , \quad [K K^a] \, , \quad [KK] 
\label{spin2comp}
\end{align}
for spin 2 currents and
\begin{align}
[J^a G^\pm] \, , \quad [K^a G^\pm] \, , \quad  [K G^\pm] \, , \quad [J^a G^{\pm ,b}] \, , \quad [K^a G^{\pm ,b}] \, , \quad [K G^{\pm ,a}] 
\end{align}
for spin 5/2 currents.
The composite spin 3 currents are
\begin{align}
\begin{aligned}
&[J^{(a }J^{b} J^{c)}] \, , \quad  [K^{a} J^{(b} J^{c)}]  \, , \quad   [J^{a}K^{(b} K^{c)}]\, , \quad  [K^{(a }K^{b} K^{c)}]\, , \quad  [KJ^{(a} J^{b)}]\, ,\\  
&  [KK^{(a} K^{b)}]  \, , \quad [K J^a K^b] \, , \quad [KKJ^a]  \, ,  \quad [KKK^a] \, , \quad [KKK] \, , \quad
[K { K^a } ' ] \, , \quad [K {J^a} '] \, , \\ 
& [J^{[a} {J^{b]}}' ]\, , \quad  [J^{a} {K^{b}}' ] \, , \quad [K^{[a} {K^{b]}}' ] \, , \quad 
 [J^a Q^b] \, , \quad [K^a Q^b] \, ,   \quad [K Q^a] \, , \\
&[G^- G^+] \, , \quad [G^\pm G^{\mp,a}] \, , \quad [G^{+,a} G^{-,b} ] \, , \quad [G^\pm G^{\pm ,a}] \, , \quad [G^{\pm , [a} G^{\pm , b]}] \, .
\end{aligned}
\end{align}
We can check that the number of independent primary operators with conformal weight $i/2$ matches with $d(i/2)$ in \eqref{dh}.

\subsubsection{Associativity of OPE }

In the following, we mainly consider the cases with $M=2,3,4$.
\footnote{With $M=2$, the OPEs among generators were analyzed by fully making use of the large $\mathcal{N}=4$ superconformal symmetry \cite{Beccaria:2014jra}. With $M>2$, there is no such supersymmetry, and the same method cannot be applied.}
We expand the operator product $\Phi^{(s_1)} \times \Phi^{(s_2)}$ by (composite) operators up to spin $s_1 +s_2 -1$, where we collectively denote the spin $s$ operators by $\Phi^{(s)}$.
The coefficients depend on the $su(M)$ indices and we express them by making use of invariant tensors in appendix \ref{sec:tensors}, which are simplified  for $M=2$ as in  \eqref{d4}, \eqref{d5}, \eqref{d6}.
We start by examining the OPEs of $\Phi^{(s_1)} \times \Phi^{(s_2)}$ with the smallest $s_1+s_2$, then move to the cases with larger $s_1 + s_2$. 
The smallest cases are with $s_1 = s_2 = 1$, but we already knew these OPEs as in \eqref{KKOPE} and \eqref{JJKKOPE}.

The simplest non-trivial OPEs are for $\Phi^{(1)} \times \Phi^{(3/2)}$.
Examining the associativity of  $\Phi^{(1)} \times \Phi^{(1)} \times \Phi^{(3/2)}$, we can fix the OPEs uniquely up to the overall normalizations of currents. For the OPEs involving $K$, we can set
\begin{align}
K(z) G^\pm (0) \sim \pm \frac{ G^\pm (0)}{z} \, , \quad K (z) G^{\pm ,a} (0) \sim \pm \frac{ G^{\pm ,a} (0)}{z}
\label{KGOPE}
\end{align}
by properly redefining $K$. In particular, the above OPEs fix the constant $\kappa_K$ in \eqref{KKOPE} as seen below.
The OPEs between $J^a,K^a$ and $G^{\pm}$ can be regarded as definitions of $G^{\pm,a}$, and our choice is
\begin{align}
J^a (z) G^\pm (0) \sim \pm \frac{G^{\pm , a} (0)}{z} \, , \quad
K^a (z) G^\pm (0) \sim\mp \frac{G^{\pm , a} (0)}{z} \, . \label{JKG0OPE}
\end{align}
With these definitions, the OPEs among $J^a,K^a$ and $G^{\pm,a}$ are 
\begin{align}
\begin{aligned}
J^a (z) G^{\pm, a} (0) \sim \pm \frac{1}{M} \delta^{ab} \frac{G^{\pm}}{z} 
+ \left ( \frac{i}{2} f^{ab}_{~~c} \mp \frac{1}{2} d^{ab}_{~~c} \right ) \frac{G^{\pm ,a}}{z} \, , \\
K^a (z) G^{\pm, a} (0) \sim \mp \frac{1}{M} \delta^{ab} \frac{G^{\pm}}{z} 
+ \left ( \frac{i}{2} f^{ab}_{~~c} \pm \frac{1}{2} d^{ab}_{~~c} \right ) \frac{G^{\pm ,a}}{z} \, .
\end{aligned} \label{JKGaOPE}
\end{align}
There is an extra freedom to choose the relative sign in front of $d^{ab}_{~~c}$, and we have chosen one of them.

We  then examine the OPEs of $\Phi^{(1)} \times Q^a$ from the associativity of $\Phi^{(1)} \times \Phi^{(1)} \times Q^a$.
The OPEs of $\Phi^{(1)} \times Q^a$ are schematically written as
\begin{align}
J^a \times Q^b \sim \, & \delta^{ab} (a_{11} K + a_{12} [KK] ) \nonumber \\
& + i f^{ab}_{~~c} (a_{21} J^c + a_{22} K^c + a_{23} Q^c + a_{24} [K J^c] + a_{25} [K K^c])  \nonumber \\
&+ d^{ab}_{~~c} (a_{31} J^c + a_{32} K^c + a_{33} Q^c + a_{34} [K J^c] + a_{35} [K K^c])  \nonumber  \\
&+ A^{ab}_{4,cd} [J^{(c} J^{d)}] + A^{ab}_{5,cd} [K^{(c} K^{d)}] + A^{ab}_{6,cd} [J^{c} K^{d}] \, ,  \nonumber \\
K^a \times Q^b \sim \, & \delta^{ab} (b_{11} K + b_{12} [KK] ) \label{A1QOPE}
\\
& + i f^{ab}_{~~c} (b_{21} J^c + b_{22} K^c + b_{23} Q^c + b_{24} [K J^c] + b_{25} [K K^c]) \nonumber  \\
&+ d^{ab}_{~~c} (b_{31} J^c + b_{32} K^c + b_{33} Q^c + b_{34} [K J^c] + b_{35} [K K^c])  \nonumber  \\
&+ B^{ab}_{4,cd} [J^{(c} J^{d)}] + B^{ab}_{5,cd} [K^{(c} K^{d)}] + B^{ab}_{6,cd} [J^{c} K^{d}] \, , 
 \nonumber \\ K \times Q^a \sim \, & k_1 J^a + k_2 K^a + k_3 Q^a + k_4 [K J^a] + k_5 [K K^a] + i f^{a}_{~bc} k_6 [J^b K^c] \nonumber  \\
&+ d^{a}_{~bc} (k_7 [J^{(b} J^{c)}] + k_8 [K^{(b} K^{c)}] + k_9 [J^{b} K^{c}] )  \nonumber 
\end{align}
up to contributions from descendants.
As in the bosonic case, we use small letters like $a_{ij}$ for constants without indices and capital ones like $A_i$ for constants with indices. The coefficients expressed by capital ones can be expanded by invariant tensors as in \eqref{C3abcd}.

While restricting the parameters by solving the conditions from associativity of the OPEs, we may encounter several discrete choices for $a_{23},b_{23},k_3$. These values correspond to the charges of $Q^a$ w.r.t.~$J^a,K^a,K$. From the reduction of $sl(2 M|M)$, we can see that $Q^a$ are charged w.r.t.~only one of $J^a$ and $K^a$ and uncharged w.r.t.~$K$. From this, we set $a_{23} \neq 0$ and $b_{23} =  k_3 = 0$. Originally we have a symmetry under the exchange of $J^a$ and $K^a$, but this choice breaks the symmetry.

Among the parameters used in \eqref{A1QOPE}, some of them can be removed by redefining the operators.
Using the spin 2 composite primaries in \eqref{spin2comp}, we can redefine the spin 2 primary $Q^a$ by
\begin{align}
\begin{aligned}
Q^a \to& z_0 Q^a + z_1 [K J^a] + z_2 [K J^a] + i f ^{a}_{~bc} z_3 [J^{b} K^c] \\
& \qquad \qquad + d^a_{~bc} (z_4 [J^{(b} J^{c)}] + z_5 [K^{(b} K^{c)}] + z_6 [J^{b} K^{c}]) \, .
\end{aligned}
\label{Qredef}
\end{align} 
Using $z_1,z_2,z_4,z_5,z_6$, we can set the parameters $a_{11},b_{11},a_{31},a_{32},b_{32}$ to vanish. Moreover, with $z_3$, we remove one parameter in $A^{ab}_{6,cd}$. There is still one ambiguity from $z_0$, which corresponds to the overall factor of $Q^a$. 
With this definition of $Q^a$, the OPEs are drastically simplified as
\begin{align}
J^a (z) Q^b (0) \sim \frac{i f^{ab}_{~~c} Q^c}{z} \, , \quad 
K^a (z) Q^b (0) \sim 0 \, , \quad K (z)Q^a (0) \sim 0 \, .
\label{A1A2OPE}
\end{align}

The OPEs analyzed above have $s_1 + s_2 =3$, but there are other OPEs with the same $s_1 +s_2$, i.e., $\Phi^{(3/2)} \times \Phi^{(3/2)}$.
Our ansatz for the OPEs may be written as
\begin{align}
G^+ \times G^- \sim \, & c_0 I + c_1 K + c_2 [KK] + c_3 [J^a J^a] + c_4 [K^a K^a] + c_5 [J^a K^a] \, , \nonumber \\
G^+ \times G^{-,a} \sim \, & d_1 J^a + d_2 K^a + d_3 [KJ^a] + d_4 [KK^a] + d_5 Q^a + i f^{a}_{~bc} d_6 [J^b K^c] \nonumber\\
&+ d^a_{~bc} (d_7 [J^{(b} J^{c)}] + d_8 [K^{(b} K^{c)}] + d_9 [J^b K^c]) \, , \nonumber\\
G^- \times G^{+,a} \sim \, & e_1 J^a + e_2 K^a + e_3 [KJ^a] + e_4 [KK^a] + e_5 Q^a + i f^{a}_{~bc} e_6 [J^b K^c] \nonumber \\ & + d^a_{~bc} (e_7 [J^{(b} J^{c)}] + e_8 [K^{(b} K^{c)}] + e_9 [J^b K^c]) \, , \\
G^{+,a} \times G^{-,b} \sim \, & \delta^{ab} (f_{0} I + f_{11} K + f_{12} [KK] ) \nonumber\\
& + i f^{ab}_{~~c} (f_{21} J^c + f_{22} K^c + f_{23} Q^c + f_{24} [K J^c] + f_{25} [K K^c]) \nonumber\\
&+ d^{ab}_{~~c} (f_{31} J^c + f_{32} K^c + f_{33} Q^c + f_{34} [K J^c] + f_{35} [K K^c])  \nonumber\\
&+ F^{ab}_{4,cd} [J^{(c} J^{d)}] + F^{ab}_{5,cd} [K^{(c} K^{d)}] + F^{ab}_{6,cd} [J^{c} K^{d}] \, .\nonumber
\end{align}
Here we have an additional ambiguity related to the overall factor of $G^{\pm}$ along with that of $Q^a$ parametrized by $z_0$ in \eqref{Qredef}. For the corresponding parameters, we set $c_0, d_5$ arbitrary.
Solving the constraints from the associativity of $\Phi^{(1)} \times \Phi^{(3/2)} \times \Phi^{(3/2)}$, we express the parameters only in terms of $c_0,d_5,\kappa_K, \ell_1 ,\ell_2 ,c$.

We could obtain further constraints from  the associativity of $\Phi^{(3/2)} \times \Phi^{(3/2)} \times \Phi^{(3/2)}$. For this, we generically need the information of $\Phi^{(3/2)} \times Q^a$ as well.
However, the OPE $G^+ \times G^-$ does not involve $Q^a$ contrary to the other cases.
Therefore, we can examine the associativity of, say, $G^- \times G^+ \times G^-$, and we obtain
\begin{align}
\kappa_K =  \frac{ 3 \ell_1 \ell_2 M }{  \ell_1 + \ell_2 + M } \, , \quad  c = \frac{(\ell_1 +  \ell_2 ) (M^2 -1) + 3 \ell_1 \ell_2 M }{  \ell_1 + \ell_2 + M } \, .
\label{asskc}
\end{align}
With these values, the OPE $G^+ \times G^-$ is written as
\begin{align}
G^+ (z) G^- (0) \sim \frac{c_0}{z^3} +\frac{3 c_0}{2 \kappa_K}\left[ \frac{2 K }{z^2} + \frac{2 \tilde T + K ' }{z} \right]
\end{align}
with 
\begin{align}
\tilde T =   T -  \frac{1 }{ 2 ( \ell_1 + \ell_2 + M)} (( J^a + K^a)( J^a + K^a)) \, .
\label{tT}
\end{align}
In order to reduce one more parameter as expected, we need the decoupling condition of, say, $G^{\pm(5/2),a}$ for $n=1$, which will be incorporated in other OPEs, such as $G^+ \times G^{-,a}$

Similarly, we examine $\Phi^{(3/2)} \times Q^a$.
From the associativity for  $\Phi^{(1)} \times \Phi^{(3/2)} \times Q^a$ and $\Phi^{(3/2)} \times \Phi^{(3/2)} \times \Phi^{(3/2)}$, we express the OPEs of the forms $\Phi^{(3/2)} \times \Phi^{(3/2)}$ and 
$\Phi^{(3/2)} \times Q^a$ in terms of $ \ell_1$ up to the overall factors parameterized by $c_0 ,d_5 $.  In particular, we obtain
\begin{align}
\ell_2 = - \frac12 ( \ell_1 + M) \, .
\end{align}
We have checked that the central charge $c$ in \eqref{asskc} and the above $\ell_2$ in terms of $\ell_1$ are consistent with the expressions in \eqref{cell2} for $n=1$.
For $M=2,3$, we have also examined the OPEs of $Q^a \times Q^b$ and checked that they are written only in terms of  $ \ell_1$ up to the ambiguities of $c_0 ,d_5$.

\subsection{Degenerate representations}
\label{sec:sminimal}

In the previous subsection, we have obtained the OPEs among generators at $n=1$.
With the information, we examine degenerate representations of the $\mathcal{N}=2$ W-algebra.
For simplicity, we set $M=2$.
In subsection \ref{sec:null}, we have examined  degenerate representations for the bosonic case using commutation relations in terms of mode expansions. 
Instead of working with commutation relations, we examine representations mainly by making use of OPEs as in   \cite{Hornfeck:1993kp,Candu:2012tr,Candu:2012ne,Candu:2013uya}, see also appendix \ref{sec:minimal}.

In order to obtain physical intuition, we start by working with mode expansions and then move to OPE language. The mode expansions of generators are
\begin{align}
&J^a (z) = \sum_{n \in \mathbb{Z}} \frac{J^a_n}{z^{n+1}} \, , \quad K^a (z) = \sum_{n \in \mathbb{Z}} \frac{K^a_n}{z^{n+1}} \, , \quad K (z) = \sum_{n \in \mathbb{Z}} \frac{K_n}{z^{n+1}} \, , \quad  G^{\pm} (z) = \sum_{r \in \mathbb{Z} +1/2} \frac{G_r^{\pm}}{z^{r + 3/2}} \, , \nonumber \\
&G^{\pm,a} (z) = \sum_{r \in \mathbb{Z} +1/2} \frac{G_r^{\pm,a}}{z^{r + 3/2}} \, , \quad 
T(z ) = \sum_{n \in \mathbb{Z}} \frac{L_n}{z^{n +2}} \, , \quad
Q^a (z) = \sum_{n \in \mathbb{Z}} \frac{Q^a_n}{z^{z+2}} \, .
\end{align}
We introduce states $|j\rangle$ with $j=1,2$ primary w.r.t.~the $\mathcal{N}=2$ W-algebra.
The zero modes of the algebra are $L_0,J^a_0,K^a_0,K_0,Q^a_0$. The first four zero-modes commute with each other, and the basis is chosen such that
\begin{align}
L_0| j \rangle = h | j \rangle \, , \quad J^a_0 | j \rangle = - ( t^a )_{j}^{~i}  |i \rangle  \, , \quad K^a_0 | j \rangle = 0  \, , \quad K | j \rangle = q | j \rangle \, .
\end{align}
We consider primary states  in the spin $1/2$ representation of $J^a_0$ and in the trivial representation of $K^a_0$.
In terms of the operator $\mathcal{O}_j$ corresponding to $|j\rangle$, we require the OPEs
\begin{align}
\begin{aligned}
&T(z) \mathcal{O}_j (0) \sim \frac{h \mathcal{O}_j (0)}{z^2} + \frac{{ \mathcal{O}_j } ' (0)}{z} \, , \quad J^a(z) \mathcal{O}_j (0) \sim - \frac{( t^a )_{j}^{~i} \mathcal{O}_i (0)  }{z} \, , \\
&K^a(z) \mathcal{O}_j (0) \sim 0 \, , \quad 
K(z) \mathcal{O}_j (0) \sim \frac{q \mathcal{O}_j (0)}{z} \, .
\end{aligned}
\end{align}
There is another set of zero modes $Q^a_0$.
Since they do not commute with some other sets of zero-modes, we cannot take the basis in a representation of $Q_0^a$ simultaneously. Even with this fact, we require
\begin{align}
Q^a_0 | j \rangle = w_0 (t^a)_j^{~i} | i \rangle
\label{assumption1}
\end{align}
as in \eqref{assumption0} for the bosonic case.

In the $\mathcal{N}=2$ W-algebra, there are $2 M^2 \,(=8)$ supercharges, and thus there are 8 superpartners of the primary state $|j\rangle$. In a usual $\mathcal{N}=2$ superconformal field theory, it is convenient to consider a subsector by assigning the chiral primary condition 
\begin{align}
G^{+}_{-1/2}|j \rangle = 0 \, . 
\label{chiralprimary}
\end{align}
Our $\mathcal{N}=2$ W-algebra includes a $\mathcal{N}=2$ superconformal algebra as a subalgebra, but
it is generated by $\{ \tilde T , K , G^\pm \}$ with the modified energy-momentum tensor $\tilde T$ defined in \eqref{tT}.
We assign the chiral primary condition \eqref{chiralprimary} as in the usual $\mathcal{N}=2$ theory.
From $ \langle  j | G^{-}_{1/2} G^+_{-1/2} | j \rangle = 0 $, we have
\begin{align}
2 h - \frac{3}{2 + k} = q \, .
\end{align}
There is a shift of conformal weight due to the modification of energy-momentum tensor.
We further observe that
\begin{align}
G^{+,a}_{- 1/2} | j \rangle =  [J^a_0 , G^{+}_{- 1/2}]  | j \rangle = 0 \, ,
\end{align}
where we have used the commutation relations read off from \eqref{JKG0OPE}.
Therefore, the chiral primary condition \eqref{chiralprimary} reduces the number of superpartners by half,
and only the actions of $G^-_{-1/2}$ and $G^{-,a}_{-1/2}$ produce new states.
In terms of OPEs, we introduce corresponding new operators by
\begin{align}
G^- (z) \mathcal{O}_j (0) \sim \frac{\tilde{\mathcal{O}}_j (0)}{z} \, , \quad
G^{-,a} (z) \mathcal{O}_j (0) \sim \frac{\tilde{\mathcal{O}}^{a}_{j} (0)}{z} \, .
\end{align}
The operators $\tilde{\mathcal{O}}_j $ and $\tilde{\mathcal{O}}^{a}_{j}$ have the conformal dimension $h+1/2$ and the eigenvalue of $K_0$ as $q -1$.

As in the bosonic case, we look for null states of the form $Q_{-1}^a |j\rangle + \cdots$, which means that the action of $Q_{-1}^a$ is written in terms of the other modes, such as, $J^a_{-1}, K^a_{-1}, K_{-1}, L_{-1}$. There could be terms like
\begin{align}
G^{+,a}_{-1/2} G^{-,b}_{-1/2}|j\rangle = \{ G^{+,a}_{-1/2} , G^{-,b}_{-1/2}\} |j\rangle \, , 
\end{align}
but they can be rewritten in terms of $J^a_{-1}, K^a_{-1}, K_{-1}, L_{-1}$.
From the form of null states along with \eqref{assumption1}, we use the ansatz as
\begin{align}
\begin{aligned}
&Q^a (z) \mathcal{O}_j (0) \sim  (t^a)_j^{~i}  \left[ \frac{w_0 \mathcal{O}_i (0)}{z^2} + \frac{w_1 {\mathcal{O}_i}'  (0) + w_2 (K \mathcal{O}_i)(0)}{z} \right]  \\
& \qquad  + \frac{w_3 (J^a \mathcal{O}_j) (0)+ w_4 (K^a \mathcal{O}_j) (0)}{z} + i f^{a}_{~bc} (t^b)_j^{~i}  \left[ \frac{w_5 (J^c \mathcal{O}_i) (0)+ w_6 (K^c \mathcal{O}_i) (0)}{z}\right] \, .
\end{aligned} \label{QOOPE}
\end{align}
As argued in \cite{Hornfeck:1993kp},
the primary conditions for the null vectors are expected to be examined by the OPE associativity involving $Q^a \times \mathcal{O}_j$.
From the associativity of $\Phi^{(1)} \times Q^a \times \mathcal{O}_j$, we can fix the coefficients except for $w_0$.

In order to fix $w_0$ as well as the conformal weight $h$, we shall examine the associativity of  $Q^a \times Q^b \times \mathcal{O}_j$. Before doing so, we need to examine the OPEs involving $\tilde {\mathcal{O}_j}$ and $\tilde{\mathcal{O}_j^a}$, since the operator product $Q^a \times Q^b$ produces  composite operators involving $G^{\pm , A}$.
We first study the OPEs of the forms $\Phi^{(1)} \times \tilde{\mathcal{O}}_j$ and $\Phi^{(1)} \times \tilde{\mathcal{O}}^{a}_{j}$ by examining the associativity for  $\Phi^{(1)} \times \Phi^{(1)} \times  \tilde{\mathcal{O}}_{j}$, $\Phi^{(1)} \times \Phi^{(1)} \times \tilde{\mathcal{O}}^{a}_{j}$, and
$\Phi^{(1)} \times G^{-,a} \times \mathcal{O}_{j}$. We can fix the OPEs as
\begin{align}
\begin{aligned}
&J^a (z) \tilde{\mathcal{O}}_{j} (0) \sim - \frac{  ( t^a )_{j }^{~i} \tilde{\mathcal{O}}_{i} (0)   + \tilde{\mathcal{O}}^{a}_{j} (0)}{z} \, ,  \\
&J^a (z) \tilde{\mathcal{O}}^{b}_{j} (0) \sim \frac{ - \frac14 \delta^{ab}\tilde{\mathcal{O}}_{j} (0) + \frac{i}{2} f^{ab}_{~~c} \tilde{\mathcal{O}}^{c}_{j} -  ( t^a )_{j}^{~i} \tilde{\mathcal{O}}^{b}_{i}  }{z} \, , \\
&
K^a (z) \tilde{\mathcal{O}}_{j} (0) \sim  \frac{ \tilde{\mathcal{O}}^{a}_{j} (0)}{z} \, , \quad K^a (z) \tilde{\mathcal{O}}^{b}_{j} (0) \sim \frac{  \frac14 \delta^{ab}\tilde{\mathcal{O}}_{j} (0) + \frac{i}{2} f^{ab}_{~~c} \tilde{\mathcal{O}}^{c}_{j} (0) }{z} \, . 
\end{aligned}
\end{align}
We next examine the OPEs of 
\begin{align}
G^+  \times \tilde{\mathcal{O}}^j \, , \quad G^+  \times \tilde{\mathcal{O}}^{a,j} \, , \quad
G^{+,a} \times \tilde{\mathcal{O}}^j \, , \quad G^{+,a} \times\tilde{\mathcal{O}}^{b,j} \, , \label{CPOPE1}
\end{align}
which can be written as linear combinations of
\begin{align}
[\mathcal{O}^j] \, , \quad  [K \mathcal{O}^j] \, ,\quad  [J^a \mathcal{O}^j] \, ,\quad  [K^a \mathcal{O}^j] \, . \label{CPOPE2}
\end{align}
We provide our ansatz for these OPEs in appendix \ref{sec:opes}.
The coefficients can be determined from the associativity of $\Phi^{(1)} \times \Phi^{(3/2)} \times \tilde{\mathcal{O}}_j$ and $\Phi^{(1)} \times \Phi^{(3/2)} \times \tilde{\mathcal{O}}^{a}_{j}$.
Since $G^+ \times G^-$ does not generate $Q^a$, we can also utilize the associativity of $G^+ \times G^- \times \mathcal{O}_j$.

With the above preparations, we can examine the associativity of $Q^a \times Q^b \times \mathcal{O}_j$ with the ansatz \eqref{QOOPE}.
From this, $w_0$ can be fixed as%
\footnote{As in the bosonic case analyzed in appendix \ref{sec:minimal}, we have used only a part of conditions coming from associativity of  the OPEs. We believe that every condition from the associativity is satisfied up to null vectors with \eqref{w0value} and \eqref{candconf}. However, we have not checked it yet.}
\begin{align}
w_0 = -\frac{c_0 (2 h (\ell_1+2)+(\ell_1-1) (\ell_1+3))}{4 d_7  \ell_1^2 (\ell_1+2)}  \, , \label{w0value}
\end{align}
and $h$ should be one of
\begin{align}
h =  \frac{1-\ell_1}{2 \ell_1 +4} \, , \quad h =  \frac{\ell_1+5}{2 \ell_1+4}- \ell_1 \, .
\label{candconf}
\end{align}
These results resemble those of the bosonic case given in \eqref{h00}.
We will see below that the first conformal weight can be realized for a state of the coset \eqref{scoset}.

\subsection{Dual coset CFT}
\label{sec:sdualcoset}

As argued before, the $\mathcal{N}=2$ W-algebra with $su(M)$ symmetry can be realized as the asymptotic symmetry of the $\mathcal{N}=2$ higher spin supergravity of \cite{Prokushkin:1998bq} with $M \times M$ matrix valued fields.
On the other hand, it was proposed  in \cite{Creutzig:2013tja} that the classical higher spin sugergravity is dual to the coset \eqref{scoset} at a large $N$ limit.%
\footnote{See appendix \ref{sec:alternatives} for an alternative proposal of dual coset.}
The central charge of the coset is
\begin{align}
c = M^2 + 3 M N -1  - \frac{M^3 + 3 M N^2 + 3 N M^2 - M}{k + N + M} \, .
\end{align}
From the holography, we would like to claim that  the $\mathcal{N}=2$ W-algebra can be realized as the symmetry algebra of the coset \eqref{scoset} even with finite $N$. 
In the following, we collect strong evidence supporting the claim.

The model has the symmetry of two affine $su(M)$ algebras. One of them comes from $su(M)_k \subset su(N+M)_k$ in the numerator. The other is $su(M)_N$ constructed from $su(N)$ invariant combinations of $NM$ complex fermions from $so(2NM)_1$. 
We first require $\ell_1 = k$ and $\ell_2 = N$. Then the match of the central charge is realized at $\lambda = - n$, where the 't Hooft parameter is defined as
\begin{align}
\lambda = \frac{N}{k + N + M} \, . \label{superthooft}
\end{align}
We can check that the map of parameters is consistent with the expressions of $c$ and $\ell_2$ in \eqref{cell2}.
Using the symmetry under the exchange of two $su(M)$ currents, 
we may require $\ell_1 = N$ and $\ell_2 = k$. Then the correspondence happens at $\lambda = - n$ with
\begin{align}
\lambda = \frac{k}{k + N + M}  \, ,
\end{align}
where $N$ and $k$ are exchanged. Thus, there are two ways to realize the $\mathcal{N}=2$ W-algebra in terms of the coset \eqref{scoset}, and this indicates the existence of duality for the coset \eqref{scoset}.
We will come back to this point later.

\subsubsection{Symmetry generators}

As in the bosonic case, we explicitly construct the low spin generators of the $\mathcal{N}=2$  W-algebra in terms of the coset \eqref{scoset}. See \cite{Ahn:2013oya,Gaberdiel:2014yla} for the case with $M=2$.
For this, we adopt the same notation as in the bosonic case. We decompose $su(N+M)$ in the numerator as in \eqref{suNMdec}. We use the generators $t^A = (t^\alpha , t^a , t^{u(1)} , t^{( \rho \bar \imath ) } , t^{ (  \bar \rho  i ) })$ with  the metric $g^{AB} = \text{tr} (t^A t^B)$ in \eqref{suNMmetric} and the invariant tensors in \eqref{suNMtensor} with \eqref{exptensor}. The $su(M)$ currents $J^A$ and the complex fermions $(\psi^{( \rho \bar \imath ) } , \psi^{(\bar \rho i) } )$ satisfy \eqref{suNMJJOPE} and
\begin{align}
\psi^{(\rho \bar \imath )} (z)  \psi^{(\bar \rho i)} (0)  \sim \frac{\delta^{\bar \rho \rho} \delta^{i \bar \imath}}{z} \, .
\end{align}
In order to construct the symmetry generators in terms of coset \eqref{scoset}, we introduce $su(N) \oplus su(M) \oplus u(1)$ currents from $so(2 N M)_1$ as
\begin{align}
\begin{aligned}
&J_f^\alpha =  (  \psi^{( \rho \bar \imath )} \psi^{( \bar \sigma i )} )  ( t^\alpha )_{\rho \bar \sigma} \delta_{\bar \imath i} \, , \quad
J_f^a =  - ( \psi^{( \rho \bar \imath )}  \psi^{( \bar \rho j )} )  ( t^a )_{j \bar \imath} \delta_{\bar \rho \rho} \, , \\ 
&J_f^{u(1)} = ( \psi^{(\rho \bar \imath)} \psi^{(\bar \rho i)} ) \delta_{\rho \bar \rho} \delta_{\bar \imath i} \, .
\label{Jfs}
\end{aligned}
\end{align}
In particular, the currents in the denominator of the coset \eqref{scoset} are given by
\begin{align}
\tilde J^\alpha = J^\alpha + J_f^\alpha \, , \quad
\tilde J^{u(1)} = \sqrt{MN (N + M)} J^{u(1)} + (N +M) J_f^{u(1)} \, .
\label{tJs}
\end{align}

We start by constructing spin 1 currents in the $\mathcal{N}=2$ W-algebra.
One of them is simply given by $J^a$ and another is $K^a = J_f^a$ defined in \eqref{Jfs}.
The other spin 1 current $K$ is given by a linear combination of $J^{u(1)}$ and $J_f^{u(1)}$ and should be regular w.r.t.~$\tilde J^{u(1)}$ in \eqref{tJs}.  We identify $K$ by
\begin{align}
K = \frac{1}{N + M + k} \left( \sqrt{M N (N + M)} J^{u(1)} - k J_f^{u(1)} \right) \, ,
\end{align}
where the overall normalization is chosen such as to reproduce \eqref{KKOPE}.
For spin 3/2 currents, we can construct them from products of $ \psi^{(\sigma \bar \jmath)}$  and  $J^{(\bar \rho i)}$
 or those of  $\psi^{(\bar \sigma j)}$ and  $J^{(\rho \bar \imath)}$.
Taking care of the properties under the $su(M)$ action, we find
\begin{align}
\begin{aligned}
&G^- = (  \psi^{(\rho \bar \imath)}  J^{(\bar \rho i)})  \delta_{\rho \bar \rho} \delta_{\bar \imath i} \, , \quad
G^+ =  (   J^{(\rho \bar \imath)}  \psi^{(\bar \rho i)} ) \delta_{\rho \bar \rho} \delta_{\bar \imath i} \, , \\
&G^{-,a} = - ( \psi^{(\rho \bar \imath)} J^{(\bar \rho j)}) (t^a)_{j \bar \imath} \delta_{\rho \bar \rho} \, , \quad G^{+,a} = - (J^{(\rho \bar \imath)} \psi^{(\bar \rho j)}) (t^a)_{j \bar \imath} \delta_{\rho \bar \rho} \, ,
\end{aligned}
\end{align}
where the overall factors are set to reproduce the OPEs \eqref{JKG0OPE} and \eqref{JKGaOPE}.
A spin 2 current is given by the energy-momentum tensor, which can be constructed by the standard coset construction \cite{Goddard:1984vk}.
For charged spin 2 currents, we require the OPEs \eqref{A1A2OPE} along with the condition primary w.r.t.~the Virasoro algebra. Starting from all possible linear combinations of products of spin 1 currents in the coset \eqref{scoset}, we find the expressions of $Q^a$ as
\begin{align}
\begin{aligned}
Q^a =& [(J^{(\rho \bar \imath)} J^{( \bar \rho j )}) +( J^{( \bar \rho j )}J^{(\rho \bar \imath)}) ] \delta_{\rho \bar \rho} (t^a)_{j \bar \imath}\\ & - \frac{N}{M + 2k} d^{a}_{~bc} (J^b J^c) 
+ \frac{2}{k} \sqrt{\frac{N (N + M)}{M}} ( J^a J^{u(1)}) \, ,
\end{aligned}
\end{align}
which are exactly the same as \eqref{Qcoset} for the bosonic case.
This is because we redefined the spin 2 currents $Q^a$ such that the OPEs become the same as the bosonic ones.  We have checked that the OPEs among generators reproduce the previous ones up to null vectors for several explicit examples.

\subsubsection{Spectrum}

In the previous subsection, we studied the degenerate representations of the $\mathcal{N}=2$ W-algebra, and in particular, we obtained the conformal weights \eqref{candconf} of primary states belonging to degenerate representations. Here we compare the primary states with those in the coset \eqref{scoset}. The state can be labeled by $(\Lambda_{N+M}, \omega ; \Lambda_N ,m)$, where $\Lambda_L$ and $m \in \mathbb{Z}_{\hat \kappa}$ denote the highest weight of $su(L)$ and $u(1)$ charge, respectively. In addition to them, we use $\omega = -1,0,1,2$ for the representation of $so(2 N M)_1$. The conformal weight of the state $(\Lambda_{N+M}, \omega ; \Lambda_N ,m)$ is
\begin{align}
h = n + h^{N+M,k}_{\Lambda_{N+M}} + h^{2 N M}_\omega - h^{N,k+M}_{\Lambda_N} - h^{\hat \kappa}_m \, , 
\end{align}
where $h^{L,K}_{\Lambda_L}$ and $h^{\hat \kappa}_m$ were defined in \eqref{cwformulasub} and $n$ is an integer related to the embedding of representations. Moreover, we have introduced $h^{2NM}_\omega = \omega/4$ for $\omega =0,2$ and $h^{2NM}_\omega = NM/8$ for $\omega = \pm 1$.
As in the bosonic case, we consider the basic states
\begin{align}
(\text{f}, 0; 0, N) \otimes (\bar{\text{f}}, 0 ; 0 , -N )\, , \quad (0,0;\text{f}, - N - M ) \otimes   (0,0;\bar{\text{f}},  N + M ) \, ,  
\end{align}
whose conformal weights are
\begin{align}
h = \frac{k \lambda M+\lambda+M^2-1}{2 M (k+M)} \, , \quad h = \frac{- \lambda M (k+M)+2 M (k+M)+\lambda-1}{2 M (k+M)}
\end{align}
in terms of the 't Hooft parameter $\lambda$ in \eqref{superthooft}.
With $\lambda = - n = - 1$ and $M=2$, the first expression becomes  
\begin{align}
h =  \frac{1-k}{2 k+4} \, .
\end{align}
Setting $k = \ell_1$, this reproduces the first expression in \eqref{candconf}.

\subsubsection{Decompositions of the symmetry algebra}

It is possible to learn some properties of the W-algebra by making use of its coset realization.
Here we would like to achieve this by decomposing the coset algebra.

As mentioned above, the W-algebra includes the $\mathcal{N}=2$ superconformal algebra with the modified energy-momentum tensor \eqref{tT}, and this fact was utilized for the analysis of degenerate representations.
We can decompose the symmetry algebra of the coset \eqref{scoset} as
\begin{align}
\frac{su(N+M)_k \oplus so(2 N M)_1}{su(N)_{k + M}\oplus u(1)_{\hat \kappa}}  \supset
\frac{su(N+M)_k \oplus so(2 N M)_1}{su(N)_{k + M} \oplus su(M)_{k+N} \oplus u(1)_{\hat \kappa}} 
\oplus su(M)_{k + M} \, .
\end{align}
The first term in the right hand side is nothing but the Grassmannian Kazama-Suzuki model \cite{Kazama:1988uz,Kazama:1988qp}.
It is constructed to have the $\mathcal{N}=2$ superconformal symmetry, and the energy-momentum tensor is given in \eqref{tT}. Such a decomposition is often useful in analyzing representations. 

In \cite{Creutzig:2018pts}, we have explained the duality of the coset \eqref{coset} by utilizing its decomposition, where the decomposition can be explained in terms of brane junctions  \cite{Gaiotto:2017euk,Creutzig:2017uxh,Prochazka:2017qum,Prochazka:2018tlo,Harada:2018bkb}.
Here we apply the arguments to  the $\mathcal{N}=2$ W-algebra.
For this, we decompose the coset algebra as
\begin{align}
\label{N=2dec}
&\frac{su(N +M)_k \oplus so(2 N M)_1}{su(N)_{k +M} \oplus u(1)} \supset
\frac{su(N+M)_k}{su(N + M -1)_k \oplus u(1)} \oplus \cdots  \oplus \frac{su(N+1)_k}{su(N)_k \oplus u(1)}  \\
&\oplus \frac{su(N)_k}{su(N)_{k+1} \oplus su(N)_1} \oplus \cdots \oplus  \frac{su(N)_{k+M-1} \oplus su(N)_1}{su(N)_{k+M} \oplus u(1)} \oplus \frac{so(2 N M)_1 \oplus (M-1) u(1)}{ M su(N)_1} \, . \nonumber
\end{align}
The last term in the right hand side consists of free bosons and fermions, which will be ignored.
Each component coset except for the last one is a realization of $W_\infty [\lambda]$ obtained as the Hamiltonian reduction of $hs[\lambda]$. The two cosets
\begin{align}
\frac{su(L)_K \oplus su(L)_1}{su(L)_{K+1}} \, , \quad \frac{su(K+1)_L}{su(K)_L \oplus u(1)}
\end{align} 
are known to realize the same principal W-algebra as its symmetry \cite{Arakawa2019}.
Denoting the algebra by $W_{L,K}$, the decomposition \eqref{N=2dec} is rewritten as
\begin{align}
\frac{su(N +M)_k \oplus so(2 N M)_1}{su(N)_{k +M} \oplus u(1)} \supset
 W_{N+M-1,k}\oplus \cdots \oplus  W_{N,k}  \oplus W_{k,N} \oplus \cdots \oplus W_{k+M-1,N} 
\end{align}
up to free bosons and fermions. For $M=1$, the decomposition reduces to the one discussed for the $\mathcal{N}=2$ $W_\infty [\lambda]$ in \cite{Gaberdiel:2017hcn,Gaberdiel:2018nbs,Li:2019nna}.
There is  the triality relation of \cite{Gaberdiel:2012ku} for each W-algebra appearing in the decomposition.
However, the coset \eqref{scoset} includes extra symmetry generators which connect two neighboring component cosets, see \cite{Prochazka:2017qum} for more details.
Requiring the existence of the extra symmetry generators, only $\mathbb{Z}_2$ symmetry  reversing the order of component cosets survives for $M >1$, see \cite{Creutzig:2018pts} for the bosonic case.
This $\mathbb{Z}_2$ action exchanges $N$ and $k$, and this is consistent with the duality of the coset \eqref{scoset}.

\section{Conclusion and discussions}
\label{sec:conclusion}

We studied the rectangular W-algebra with $su(M)$ symmetry, which is obtained as quantum Hamiltonian reduction of $sl(Mn)$. We decompose $sl(Mn)$ as in \eqref{slMn} and use the $sl(2)$ principally embedded in $\mathbbm{1}_M \otimes sl(n)$.  The algebra can be identified with the asymptotic symmetry of 3d higher spin gravity with $M \times M$ matrix valued fields. The matrix extension is expected to be useful to examine superstring theory from higher spin gravity including higher Regge trajectories as well.
In our previous work \cite{Creutzig:2018pts}, we examined the basic properties of the W-algebras, such as the spin content, the central charge, and the level of the $su(M)$ currents. 
Furthermore, we computed the OPEs among generators with $n=2$ and claimed that the W-algebra can be realized by the coset \eqref{coset} at $\lambda = n$ with $\lambda$ defined in \eqref{tHooft}. 
In this paper, we extended the analysis in several ways.

We first reviewed the works in \cite{Creutzig:2018pts} but slightly extended the OPE analysis by working with $n \neq 2$ but only among low spin generators and with $M=2$. We found that the OPEs are uniquely fixed by one parameter, say, the level $\ell$ of $su(M)$ currents.
We expect that there is a family of W-algebra, which may be denoted as $W^M_\infty [\ell,\lambda]$.
The algebra may be obtained as a Hamiltonian reduction of $hs_M[\lambda]$. 
It has two continuous parameters $\ell,\lambda$ with $M$ fixed as the rank of $su(M)$.
Our claim here is that the algebra can be truncated to our rectangular W-algebra at $\lambda =n$ just as 
$hs_M[\lambda]$ can be truncated to $sl(Mn)$.  This claim was justified for $n=2$ (and $M=2$) only, and it is desired to confirm for general $n$. We could truncate $W^M_\infty [\ell,\lambda]$ when the corresponding coset \eqref{coset} has integer parameters $k,N$, and the truncation should be different from the one at $\lambda = n$, see also \cite{Gaberdiel:2013vva,Gaberdiel:2014yla,Beccaria:2014jra}.
It is important to understand the nature of the truncations of $W^M_\infty [\ell,\lambda]$ furthermore.

We then investigated the degenerate representations  of the W-algebra but with $M=n=2$.
We explicitly constructed null vectors by examining the condition primary w.r.t.~the W-algebra at low levels.
One crucial assumption is that primary states are eigenstates of $Q_0^a$ along with $L_0$ and $J^a_0$ as in \eqref{assumption0}.
We would like to see what would happen if we relax the assumption.
We also obtained representations by deducing those of $sl(4)$. 
However we have not examined their properties from the viewpoints of the W-algebra. In particular, we would like to know what kind of null vectors are related to the representations generically. We then compared the results with the spectrum of the coset \eqref{coset} and the mass of conical defect geometry of the higher spin gravity constructed in \cite{Castro:2011iw}.
In particular, the conformal weights in \eqref{cw4} give the information of quantum corrections to the masses of conical defects, and it is an important problem to reproduce them from the gravity theory, see \cite{Raeymaekers:2014kea,Campoleoni:2017xyl,Hikida:2018eih}.
It is also desired to have a more systematic understanding of degenerate representations including generic $M$ and $n$.

We also examined the $\mathcal{N}=2$ rectangular W-algebras with $su(M)$ symmetry obtained as the Hamiltonian reduction of $sl(M (n+1) | Mn)$.
Here $sl(M (n+1) | Mn)$ is decomposed as in \eqref{decssl}, and the $osp(1|2)$ principally embedded in $\mathbbm{1}_M \otimes sl(n+1|n)$ is used.
We first studied the basis properties of the $\mathcal{N}=2$ W-algebras, such as, the spin content, the central charge, and the levels of two sets of $su(M)$ currents. We then fixed the OPEs among generators with a parameter $\ell_1$ for the level of a set of $su(M)$ currents but with the restriction of $n=1$. In the case of $M=2$, the restriction can be removed as in \cite{Beccaria:2014jra} by making use of the large $\mathcal{N}=4$ symmetry.  
We further claimed that the  $\mathcal{N}=2$ W-algebra can be realized by the symmetry algebra of the coset \eqref{scoset} with setting \eqref{smap1} or \eqref{smap2}.
We studied the representations with level 1 null vectors and primary states in the spin $1/2$ representation of $J_0^a$ for $M=2$.
It is worthwhile extending the analysis as was done in the bosonic case.

In order to see the relation to superstring theory, it is useful to extend the current analysis by introducing more extended supersymmetry as in \cite{Creutzig:2014ula,Hikida:2015nfa,Creutzig:2015hta}.
It was proposed that the coset model \eqref{scoset} with a critical level $k = N +M$ is dual to a version of Prokushkin-Vasiliev theory with more extended supersymmetry. 
We may further assign an invariant condition to the matrix degrees of freedom in the higher spin theory, since a closed string does not have such degrees.  In this case, a multi-particle state in the higher spin theory would correspond to a single closed string state.
It was claimed that the restricted version is dual to the Grassmannian Kazama-Suzuki model \cite{Kazama:1988uz,Kazama:1988qp}
\begin{align}
\frac{su(N+M)_{N+M} \oplus so(2NM)_1}{su(N)_{N+2 M} \oplus su(M)_{M + 2N} \oplus u(1)_{\hat \kappa}} 
\end{align}
at the critical level. 
In particular, it was found that the critical level model possesses  $\mathcal{N}=3$ enhanced supersymmetry, which enables us to discuss the relation to superstrings on AdS$_3$, see \cite{Creutzig:2014ula,Hikida:2015nfa,Creutzig:2015hta} for more details.

In this paper, we examined the rectangular W-algebra with $su(M)$ symmetry, but we can also construct other rectangular W-algebras. There are generalizations of higher spin (super)gravity  by restricting the extra matrix degrees of freedom as in \cite{Prokushkin:1998bq}, see also \cite{Eberhardt:2018plx}.
The asymptotic symmetries of these higher spin (super)gravities were given by rectangular W-algebras with $so(M)$ or $sp(2M)$ symmetry, and the analysis of \cite{Creutzig:2018pts} was applied to these algebras in \cite{Creutzig:2019wfe}.
Without the matrix extensions, higher spin holographies were proposed in \cite{Ahn:2011pv,Gaberdiel:2011nt} for the bosonic case and in \cite{Creutzig:2012ar} for the $\mathcal{N}=1$ supersymmetric case.
In this paper, we have extended the analysis of \cite{Creutzig:2018pts} in several ways.
It is interesting to apply the current extensions to the W-algebras with $so(M)$ or $sp(2M)$ symmetry as well.

\subsection*{Acknowledgements}

We are grateful to Wei Li, Cheng Peng and Takahiro Uetoko for useful discussions.
YH thanks the organizers of ESI Programme and Workshop ``Higher spins and holography'' at the Erwin Schr{\"o}dinger Institute in Vienna, where a part of this work was done.
The work of TC is supported by NSERC grant number RES0019997.
The work of YH is supported by JSPS KAKENHI Grant Number 16H02182 and 19H01896.

\appendix

\section{Commutation relations}
\label{sec:CR}

We consider the rectangular W-algebra with $M=n=2$. 
The algebra is labeled by the level $\ell$ of $su(2)$ currents. 
The central charge of the algebra is (see \eqref{clrel})
\begin{align}
c= - \frac{8 (\ell^2 -1) }{\ell + 4} + 2 \ell - 1 \, .
\end{align}
In terms of mode expansions, generators are $L_n$, $Q_n^a$, $J_n^a$ $(a=1,2,3)$ with $n \in \mathbb{Z}$. From the OPEs \eqref{OPEJJ}, \eqref{OPETJ} and \eqref{OPEQP}, we obtain \eqref{CRmodes}.
With $n=2$, the OPEs of $Q^a \times Q^b$ were obtained in \cite{Creutzig:2018pts} (and reproduced in subsection \ref{sec:OPE} for $M=2$). 
The corresponding commutation relations can be computed as
\begin{align}
&{[}Q^a _ m , Q^b _ n] = \delta^{ab} \left\{ \frac{c_1}{12} (m^3 - m) \delta_{m+n,0}  + \frac{c_2}{2} (m - n) L_{m+n}\right\} \label{QQmodes} \\ &+ i f^{ab}_{~~c} \left\{ \left( - \frac{c_3}{2} (m +1) (n+1) + c_4 (m+n+2)(m+n+1)  \right) J^c_{m+n} + c_5 (T J^c)_{m + n}  \right\} \nonumber  \\ & + \frac{(m -n)}{2} (c_6 d_{4SS1}^{abcd} + c_7 d_{4SS2}^{abcd} ) (J_{(c} J_{d)})_{m+n} + c_8 d_{4AA}^{abcd} (J_{[c} J_{d]} {}')_{m+n}   + c_9 d_{5AS}^{abcde} (J_{(c} J_d J_{e)})_{m+n} \, . \nonumber  
\end{align}
The coefficients are given by
\begin{align}
&c_2=\frac{2 c_{1} (\ell+1)}{c (\ell+2)-3 \ell}\, , \quad  c_ 3=\frac{ c_ 1}{2 \ell} \, ,  \nonumber \\  
&  c_4=\frac{c_1 (c (\ell (3 \ell (\ell+3)+2)-24)-5 \ell (3 \ell (3 \ell+7)-16)+80)}{12 \ell \left(3 \ell^2+\ell-4\right) (c (\ell+2)-3 \ell)}  \, ,  \quad c_5 =\frac{2 c_1 (\ell+1)}{\ell (c (\ell+2)-3 \ell)}\, , \nonumber \\  
&   c_6=\frac{ c_ 1 (c-2 \ell+1)}{2 (\ell-1) (c (\ell+2)-3 \ell)}  \, , \quad  c_7=\frac{ c_ 1}{2 \ell-2 \ell^2} \, , \\  
&  c_8=\frac{ c_ 1 (-c (\ell-2) (\ell+4)+\ell (27 \ell-16)-16)}{2 \ell \left(3 \ell^2+\ell-4\right) (c (\ell+2)-3 \ell)}  \, , \quad  c_9=\frac{ c_1 (\ell (c-6 \ell+1)+4)}{2 \ell \left(3 \ell^2+\ell-4\right) (c (\ell+2)-3 \ell)}  \nonumber 
\end{align}
and $d_{4AA}^{abcd},d_{4SS1}^{abcd},d_{4SS2}^{abcd},d_{5AS}^{abcde}$ are defined in \eqref{d4} and \eqref{d5}.
We have used
\begin{align}
\begin{aligned}
&(TJ^a)_{n} = \sum_{p \leq -2 } L_p J^a_{n -p} +  \sum_{p \geq -1 } J^a_{n -p} L_p \, , \quad
(J^{(a}J^{b)})_n = \sum_{p \leq -1} J^{(a}_p J^{b)}_{n -p} +  \sum_{p \geq 0 } J^{(b}_{n -p} J^{a)}_p \, , \\
&(J^{[a} J^{b]} {}')_n = \sum_{p \leq -1} (p - n - 1) J^{[a}_p J^{b]}_{n-p} +   \sum_{p \geq 0 } (p - n - 1) J^{[b}_{n -p} J^{a]}_p \, , \\
&(J^{(a}J^b J^{c)})_n = \sum_{p \leq -1, q \leq -1} J^{(a}_p J^b_q J^{c)}_{n -p-q} +   \sum_{p \geq 0, q \leq -1} J^{(b}_q J^{c}_{n -p-q} J^{a)}_p  \\ & \qquad \qquad \qquad +\sum_{p \leq -1, q \geq 0} J^{(a}_p  J^{c}_{n -p-q} J^{b)}_q +\sum_{p \geq 0, q \geq 0} J^{(c}_{n -p-q} J^b_q  J^{a)}_p  
\end{aligned}
\end{align}
for the mode expansions of composite operators.

\section{OPEs and null vectors}
\label{sec:minimal}

In subsection \ref{sec:null}, we examined null states by making use of the commutation relations in terms of the mode expansions of generators. In this subsection, we develop an alternative way by applying associativity of the OPEs as in \cite{Hornfeck:1993kp,Candu:2012tr,Candu:2012ne,Candu:2013uya}.
For simplicity, we focus on the case where null vectors appear at  level 1 and further set $J =1/2$.

We considered primary states satisfying \eqref{primarycond} and \eqref{zeromodes}.
We denote the corresponding operator as $\mathcal{O}_j$ and require the OPEs with $T$ and $J^a$ as
\begin{align}
T(z) \mathcal{O}_j (0) \sim \frac{h \mathcal{O}_j (0)}{z^2} + \frac{\mathcal{O}_j {}' (0)}{z} \, , \quad
J^a (z) \mathcal{O}_j (0) \sim - \frac{ (t^{a})^{~i}_{j} \mathcal{O}_i (0) }{z} \, .
\end{align}
We assumed that the primary states satisfy \eqref{primarycond} and \eqref{assumption0} under the action of $Q_n^a $. In terms of OPE, we use the ansatz  as
\begin{align}
Q^a (z) \mathcal{O}_j (0) \sim w_0  (t^{a})^{~i}_{j}\left (\frac{ \mathcal{O}_i }{z^2} + \frac{ {\mathcal{O}_i }'}{h z} \right)
+ w_1 \frac{[J^a \mathcal{O}_j]}{z} +  i  f^{a}_{~bc} w_2  (t^{b})^{~i}_{j} \frac{[J^c \mathcal{O}_i] }{z} \, .
\end{align}
Here we have introduced composite primary operators of conformal weight $h+1$ as
\begin{align}
[J^a \mathcal{O}_j] = (J^a \mathcal{O}_j) + \frac{1}{2 h} (t^{a})^{~i}_{j}  { \mathcal{O}_i }'  \, .
\end{align}

We  fix the parameters by requiring associativity of the OPEs.
We use the function ``OPEJacobi'' incorporated in the Mathematica package ``OPEdefs'' \cite{Thielemans:1991uw}. Writing a OPE as
\begin{align}
A(z) B(w) = \sum_{n \leq n_\text{max}} \frac{[AB]_n (w)}{(z-w)^n} \, ,
\end{align}
the function checks whether the conditions
\begin{align}
[A[BC]_p]_q = (-1)^{|A||B|} [B[AC]_q]_p + \sum_{l > 0} \binom{q-1}{l-1} [[AB]_l C]_{p+q - l}
\end{align}
are satisfied or not  for $p,q > 0$. Here $|A|$ represents the parity of $A$.
It was argued in \cite{Thielemans:1994er} that  the associativity of $A \times B \times C$ is satisfied when the function generates zero up to null vectors.
In particular, if the generated operators are proportional to primary operators with non-vanishing two point functions, then the factors in front of the operators should be zero.

Requiring the associativity of $J^a \times Q^b \times \mathcal{O}_j$, we obtain conditions for $w_1$ and $w_2$. Solving the conditions, we rewrite  $w_1$ and $w_2$ as
\begin{align}
w_1 =  \frac{2 w_0(4 h+k-2)}{(k-1) (4 h (k+2)-3)} \, , \quad w_2 = \frac{(4 h-2) k w_0 +w_0}{(k-1) (4 h (k+2)-3)}
\end{align}
in terms of $h$ and $w_0$.
The associativity of $Q^a \times Q^b \times \mathcal{O}_j$ leads to constraint equations for $h$ and $w_0$.
The function ``OPEJacobi'' generates several operators, which should be null in order to satisfy the associativity.
We require that coefficients in front of $\mathcal{O}_j$ vanish, and these conditions lead to the conformal dimension $h$ as in \eqref{h00}. 
In this way, we reproduce the previous result by using the OPEs involving the primary operator $\mathcal{O}_j$.
There are other operators generated by ``OPEJacobi.''
We believe that these extra operators are null, but we have not checked it yet.

\section{Technical details on the $\mathcal{N}=2$ W-algebra}
\label{sec:N=2info}

In this appendix, we collect some technical materials used for computations on the $\mathcal{N}=2$ rectangular W-algebras.

\subsection{Invariant tensors}
\label{sec:tensors}

In order to expand operator products in terms of (composite) primary operators, we need the invariant tensors with several indices.
Here we list the invariant tensors used for the OPE analysis of the $\mathcal{N}=2$ W-algebra in subsection \ref{sec:sOPE}. 
The invariant tensors with 2 and 3 indices are given by \eqref{invtensors}.
With 4 indices, we use
\begin{align}
&
d_{4AA1}^{abcd} = \delta^{a}_{~[b} \delta^{c}_{~d]} \, , \quad
d_{4AA2}^{abcd} = \text{tr} (t^{[a}t^{b]} t^{[c} t^{d]}) \, , \quad
d_{4AA3}^{abcd} = \text{tr} (t^{[a}t_{[c} t^{b]} t_{d]}) \, , \\
&	d_{4SS1}^{abcd} = \delta^{ab} \delta^{cd} \, , \quad
d_{4SS2}^{abcd} = \delta^{a}_{~(c} \delta^{b}_{~d)} \, , \quad
d_{4SS3}^{abcd} = \text{tr} (t^{(a}t^{b)} t^{(c} t^{d)}) \, , \quad
d_{4SS4}^{abcd} = \text{tr} (t^{(a}t_{(c} t^{b)} t_{d)}) \, . 
\nonumber
\end{align}
We also use tensors with 5 indices $\{a,b,c,d,e\}$. We need tensors which are anti-symmetric under $a \leftrightarrow b$ and symmetric under $c \leftrightarrow b \leftrightarrow c$ as
\begin{align}
\begin{aligned}
&d_{5AS1}^{abcde} = i f^{ab(c} \delta^{de)} \, , \quad
d_{5AS2}^{abcde} = \text{tr} (t^{[a} t^{b]} t^{(c} t^d t^{e)} ) \, , \\
&d_{5AS3}^{abcde} = \text{tr} (t_{[a} t^{(c} t_{b]} t^d t^{e)} ) \, , \quad
d_{5AS4}^{abcde} = \delta^{~(c}_{[a} d_{b]}^{~de)}\, .
\end{aligned}
\end{align}
We further use tensors which are symmetric under $c \leftrightarrow d$ and anti-symmetric under $a \leftrightarrow b$, $c \leftrightarrow e$ and $d \leftrightarrow e$ as
\begin{align}
&d_{5AH1}^{abcde} = \frac{i}{2} \left \{f^{abe} \delta^{cd} - f^{ab(c} \delta^{d)e} \right\} \, ,  \nonumber\\
&d_{5AH2}^{abcde} = \frac{1}{4} \left\{ 2 \text{tr} ( t^{[a} t^{b]} t^{(c} t^{d)} t^{e} ) - (\text{tr} ( t^{[a} t^{b]} t^{(e} t^{d)} t^{c} ) - \text{tr} ( t^{[a} t^{b]} t^{(c} t^{e)} t^{d} )   )\right\} \, ,  \nonumber\\
&d_{5AH3}^{abcde} = \frac{1}{4} \left\{ 2 \text{tr} ( t^{[a} t^{b]} t^{(c} t_{e} t^{d)} ) -\text{tr} ( t^{[a} t^{b]} t^{(e} t_{c} t^{d)} ) - \text{tr} ( t^{[a} t^{b]} t^{(c} t_{d} t^{e)} )   \right\} \, ,  \nonumber\\
&d_{5AH4}^{abcde} = \frac{1}{4} \left\{ 2 \text{tr} ( t^{[a} t^{b]} t^{e} t^{(c} t^{d)} ) -\text{tr}( t^{[a} t^{b]} t^{c} t^{(e} t^{d)} ) - \text{tr} ( t^{[a} t^{b]} t^{d} t^{(c} t^{e)} )   \right\} \, , \\
&d_{5AH5}^{abcde} = \frac{1}{4} \left\{ 2 \text{tr} ( t_{[a} t^{(c} t_{b]} t^{d)} t^{e} ) -\text{tr} ( t_{[a} t^{(e} t_{b]} t^{d)} t^{c} ) - \text{tr}  ( t_{[a} t^{(c} t_{b]} t^{e)} t^{d} ) \right\} \, ,  \nonumber\\
&d_{5AH6}^{abcde} = \frac{1}{4} \left\{ 2 \text{tr} ( t_{[a} t^{(c} t_{b]} t_{e} t^{d)} ) -\text{tr} ( t_{[a} t^{(e} t_{b]} t_{c} t^{d)} )- \text{tr} ( t_{[a} t^{(c} t_{b]} t_{d} t^{e)} )\right\} \, ,  \nonumber\\
&d_{5AH7}^{abcde} = \frac{1}{4} \left\{ 2 \text{tr} ( t_{[a} t^{e} t_{b]} t^{(c} t^{d)} ) -\text{tr} ( t_{[a} t^{c} t_{b]} t^{(e} t^{d)} ) - \text{tr}  ( t_{[a} t^{d} t_{b]} t^{(c} t^{e)} ) \right\} \, , \nonumber \\
&d_{5AH8}^{abcde} = \frac{1}{4} \left\{ 2 \delta_{[a}^{~e} d^{~cd}_{b]} -\delta_{[a}^{~c} d^{~ed}_{b]} - \delta_{[a}^{~d} d^{~ce}_{b]}    \right\} \, . \nonumber
\end{align}

\subsection{OPEs involving chiral primaries}
\label{sec:opes}

In subsection \ref{sec:sminimal}, we have expanded the operator products in \eqref{CPOPE1} as linear combinations of primary operators in \eqref{CPOPE2}.
Here we write down our ansatz explicitly.
We have used the forms
\begin{align}
&G^+ (z)  \tilde{\mathcal{O}}_{j} (0) \sim \frac{m_1 \mathcal{O}_j (0) }{z^2}  + \frac{m_2 { \mathcal{O}_j  } ' (0) + m_3 (K  \mathcal{O}_j ) (0)}{z} +  ( t^a )_{j }^{~i} \frac{ m_4 (J^a \mathcal{O}_i)   (0) + m_5 (K^a \mathcal{O}_i)  (0)}{z} \, , \nonumber \\
&G^+ (z)  \tilde{\mathcal{O}}^{a}_{j} (0) \sim \frac{n_1 (J^a  \mathcal{O}_j)  (0)  + n_2 (K^a \mathcal{O}_j) (0) }{z} \label{GpOOPE}\\
&+ ( t^a )_{j}^{~i} \left[\frac{n_3 \mathcal{O}_i (0)}{z^2} + \frac{n_4 { \mathcal{O}_i } ' (0) + n_5 (K \mathcal{O}_i) (0) }{z}  \right  ]   + 
i f^{a}_{~bc}  ( t^b )_{j}^{~i} \frac{n_6 (J^c \mathcal{O}_i) (0)+ n_7 (K^c \mathcal{O}_i)(0)}{z} 
\nonumber 
\end{align}
for the OPEs involving $G^+$.
We have set the forms as
\begin{align}
&G^{+,a} (z) \tilde{\mathcal{O}}_{j} (0) \sim \frac{o_1 (J^a \mathcal{O}_j)(0) + o_2 (K^a \mathcal{O}_j)(0)}{z} \nonumber \\
&+  ( t^a )_{j}^{~i} \left[\frac{o_3 \mathcal{O}_i (0)}{z^2} + \frac{o_4 { \mathcal{O}_i } ' (0) + o_5 (K \mathcal{O}_i) (0) }{z}  \right  ]  + 
i f^{a}_{~bc} ( t^b )_{j}^{~i} \frac{o_6 (J^c \mathcal{O}_i) (0)+ o_7 (K^c \mathcal{O}_i) (0)}{z} 
\, , \nonumber  \\
&G^{+,a} (z) \tilde{\mathcal{O}}^{b}_{j} (0) \sim \delta^{ab} \left[ \frac{p_1 \mathcal{O}_j ( 0)}{z^2} + \frac{p_2 {\mathcal{O}_j} ' (0) + p_3 (K \mathcal{O}_j) (0)} {z} \right]  \nonumber  \\
&+ \delta^{ab}  (t_c)_{j}^{~i} 
\left[ \frac{p_4 (J^c \mathcal{O}_i )(0) + p_5 (K^c \mathcal{O}_i) (0)}{z} \right]
+ i f^{ab}_{~~c} \left[ \frac{p_6 (J^c \mathcal{O}_j) (0)+ p_7 (K^c \mathcal{O}_j) (0)}{z} \right] 
\label{GpaOOPE}\\ & + i  f^{ab}_{~~c}  (t^c)_{j}^{~i} \left[\frac{p_8 \mathcal{O}_i (0)}{z^2} + \frac{p_9 {\mathcal{O}_i}' (0) + p_{10} (K \mathcal{O}_i) (0)}{z} \right] +  (t^b)_j^{~i}\left[ \frac{p_{11} (J^a \mathcal{O}_i) (0)+ p_{12} (K^a \mathcal{O}_i)(0)}{z}\right]  \nonumber  \\
& +(t^a)_j^{~i} \left[ \frac{p_{13} (J^b \mathcal{O}_i) (0)+ p_{14} (K^b \mathcal{O}_i)(0)}{z}\right]  
\nonumber 
\end{align}
for the OPEs involving  $G^{+,a}$.

\section{Alternative proposals of dual coset CFTs}
\label{sec:alternative}

In \cite{Creutzig:2018pts},  we examined a coset description of rectangular W-algebras with $su(M)$ symmetry.
In this appendix, we propose an alternative coset description of the algebra by making use of superalgebra $su(N|M)$. 
Rectangular W-algebra obtained from the Hamiltonian reduction of $hs_M[\lambda]$ can be described by free bosons or fermions at the limit $\lambda \to 0$ or $\lambda \to 1$ as was constructed in \cite{Bakas:1990xu,Odake:1990rr}. The dual coset should reduce to the same free system at a large level limit, and this fast was utilized to guess the dual coset in  \cite{Creutzig:2013tja}, see also \cite{Eberhardt:2018plx}.
The coset \eqref{coset} reduces to a free boson system at the large $k$ limit. However, we may use an alternative coset which reduces to a free symplectic fermion system at the large $k$ limit. This is the idea behind the arguments in this appendix.
Furthermore, we extend the analysis by introducing the $\mathcal{N}=2$ supersymmetry.

\subsection{Rectangular W-algebras}
\label{sec:alternativeb}

Here we propose that the rectangular W-algebra with $su(M)$ symmetry can be realized by a coset
\begin{align}
	\frac{su(N|M)_k}{su(N)_k \oplus u(1)_{\kappa}}
	\label{superalgcoset}
\end{align}
with $\kappa = M N k (M - N)$ as an alternative of \eqref{coset}. Indeed the symmetry algebra has $M^2$ fields in conformal weights $1, 2, \ldots, 2n+1$ for the following reason:
The large $k$ limit of the coset reduces to the subalgebra of $NM$ symplectic fermions that is invariant under $su(N) \oplus u(1)$. This orbifold has $M^2$ fields of conformal weight $1, 2, \ldots, 2n+1$ by Theorem 4.4 of \cite{Creutzig:2016xos}.
 The type of symmetry algebra of the coset at generic level is the same as the orbifold limit by the theory of \cite{Creutzig:2012sf, Creutzig:2014lsa}. 

The problem here is to obtain the map of parameters such that the central charge $c$ and the level $\ell$ of $su(M)$ currents coincide with each other. For the level of $su(M)$, we set $\ell = - k$. 
The central charge of the model is
\begin{align}
	c=\frac{k \left(\left(N^2-1\right)+ \left(M^2-1\right)+ 1 -2 N M \right)}{k + N -M}-\frac{k \left(N^2-1\right)}{k+N}-1 \, .
\end{align}
Compared with \eqref{DSc} and \eqref{DSl}, the correspondence happens at $\lambda = n$ with
\begin{align}
	\lambda = \frac{k}{k+N} \, , \quad \lambda = - \frac{k}{k + N - M}  \, .
\end{align}
There are two choices, and this implies a duality of the coset \eqref{superalgcoset}.

We next construct the generators of rectangular W-algebra in terms of the coset \eqref{superalgcoset}.
The superalgebra $su(N|M)$ can be decomposed as
\begin{align}
	su(N|M) = su(N) \oplus su(M) \oplus u(1) \oplus (N, \bar M) \oplus (\bar N , M) \, ,
\end{align}
and the generators are denoted as $t^A = (t^\alpha ,t^a , t^{u(1)}, t^{(\rho \bar \imath)} ,t^{(\bar \rho i)} )$. Here $t^P = (t^\alpha ,t^a ,t^{u(1)})$ are Grassmann even and $t^{(\rho \bar \imath)} ,t^{(\bar \rho i)} $ are Grassmann odd. We choose the metric $g^{AB} = \text{str} (t^A t^B)$ as
\begin{align}
	\begin{aligned}
		&\text{str} (t^\alpha t^\beta) = \delta^{\alpha \beta} \, , \quad
		\text{str} (t^a t^b) = - \delta^{ab} \, , \quad
		\text{str} (t^{u(1)} t^{u(1)})  = 1 \, , \\
		&\text{str} (t^{(\rho \bar \imath)} t^{(\bar \rho i)}) = - \text{str} (t^{(\bar \rho i)} t^{(\rho \bar \imath)} ) = \delta^{\rho \bar \rho} \delta^{\bar \imath i} \, .
	\end{aligned}
\end{align}
We also introduce the structure constants as
\begin{align}
	i f^{ABC} = \text{str} ([t^A , t^B] t^C) \, , \quad
	d^{ABC} = \text{str} (\{t^A , t^B \} t^C) \, , 
\end{align}
where our convention is such that
\begin{align}
\begin{aligned}
&i f^{(\rho \bar \imath) (\bar \sigma j) u(1)} = \frac{M+N}{\sqrt{MN (M - N)}} \delta^{\bar \imath j  } \delta^{\bar \sigma \rho } \, , \quad
d^{(\rho \bar \imath) (\bar \sigma j) u(1)} = \sqrt{\frac{M-N}{MN}} \delta^{ \bar \imath  j } \delta^{\bar \sigma \rho } \, , \\
&i f^{(\rho \bar \imath) (\bar \sigma j) \alpha} = d^{(\rho \bar \imath) (\bar \sigma j) \alpha} =  (t^\alpha)^{\bar \sigma \rho}  \delta^{j \bar \imath }  \, , \quad
i f^{(\rho \bar \imath) (\bar \sigma j) a} = - d^{(\rho \bar \imath) (\bar \sigma j) a} = \delta^{\rho \bar \sigma} (t^a)^{\bar \imath j}  
\end{aligned}
\end{align}
for non-trivial expressions.
The $su(N|M)$ affine algebra consists of bosonic currents $J^P = (J^a , J^\alpha , J^{u(1)})$ and fermionic currents $J^{(\rho \bar \imath)} , J^{(\bar \rho , i)}$.
The OPEs among the $su(N|M)$ currents are now written as
\begin{align}
	\begin{aligned}
		&J^P (z) J^Q (0) \sim \frac{k g^{PQ}}{z^2} + \frac{i f^{PQ}_{~~~R}J^R (0)}{z} \, , \quad
		J^{(\rho \bar \imath)} (z) J^{(\bar \rho i)} (0) \sim \frac{k \delta^{\rho \bar \rho} \delta^{\bar \imath i}}{z^2} + \frac{d^{(\rho \bar \imath)(\bar \rho i)}_{~~~~~~P} J^P (0) }{z} \, , \\
		& J^P (z) J^{(\rho \bar \imath)} (0) \sim \frac{i f^{P (\rho \bar \imath)}_{~~~~~(\sigma \bar \jmath) } J^{(\sigma \bar \jmath)}}{z} \, , \quad J^P (z) J^{(\bar \rho i)} (0) \sim \frac{i f^{P (\bar \rho i)}_{~~~~~(\bar \sigma j) } J^{(\bar \sigma j)}}{z} \, .
	\end{aligned}
\end{align}

We construct the generators of rectangular W-algebra up to spin 2 using the $su(N|M)$ currents.
The $su(M)_{-k}$ currents are given by $J^a$.  The energy-momentum tensor can be obtained by the standard coset construction \cite{Goddard:1984vk}.  We find that 
\begin{align}
	\begin{aligned}
		Q^a =&   \left[(J^{(\rho \bar \imath)} J^{(\bar \rho i)}) -  (J^{(\bar \rho i)}J^{(\rho \bar \imath)} )  \right] \delta_{\rho \bar \rho} (t^a)_{i \bar \imath} \\
		&- \frac{N }{ M - 2 k} d^{a}_{~bc} (J^bJ^c)  - \frac{2}{k} \sqrt{\frac{N (M - N)}{M}} (J^a J^{u(1)}) 
		\label{superQa}
	\end{aligned}
\end{align}
satisfies the required OPEs with $J^a$ and $T$ as in \eqref{OPEQP}. We have checked that the OPEs of $Q^a \times Q^b$ are reproduced for several examples.

\subsection{$\mathcal{N}=2$ rectangular W-algebras}
\label{sec:alternatives}

We would like to consider a similar realization of the $\mathcal{N}=2$ W-algebra from a coset
\begin{align}
	\frac{su(N|M)_k \oplus sp(2 N M)_{-1/2}}{su(N)_{k-M}  \oplus u(1)_{\hat{\kappa}}}
	\label{superalgebrascoset}
\end{align}
with $\hat{\kappa} = N M (M - N) (k - M + N )$. 
The factor $sp(2 N M)_{-1/2}$ can be described by $NM$ pairs of symplectic bosons.
The central charge of this model is
\begin{align}
	c = \frac{k \left(\left(N^2-1\right)+\left(M^2-1\right)+1 -2  N M \right)}{k+N-M} - N M  -\frac{(k-M)\left(N^2-1\right) }{k-M+N}-1 \, .
\end{align}
The symmetry algebra includes $su(M)_{-k} \subset su(N|M)_k$ and $su(M)_{-N}$ from the symplectic bosons.
Requiring $\ell_1 = - k$ and $\ell_2 = - N$, we find the match of central charge at $\lambda =  - n$ with
\begin{align}
	\lambda = \frac{N}{k + N - M}  \, .
\end{align}
With the alternative possibility as $\ell_1 = - N$ and $\ell_2 = - k$,  the correspondence is realized at $\lambda = - n$ with
\begin{align}
	\lambda = \frac{k}{k + N - M}  \, .
\end{align}
Here we have used the expressions \eqref{sDSl} and \eqref{sDSc}.

We then construct the symmetry generators of the $\mathcal{N}=2$ W-algebra from the coset \eqref{superalgebrascoset}. We express the factor $sp(2NM)_{-1/2}$ by symplectic bosons 
$(\varphi^{(\rho \bar \imath)}, \varphi^{( \bar \rho i ) })$ satisfying
\begin{align}
	\varphi^{(\rho \bar \imath)} (z) \varphi^{(\bar \rho i)} (0) \sim \frac{\delta^{\rho \bar \rho} \delta^{\bar \imath i}}{z} \, . 
\end{align}
With the symplectic bosons, $su(N) \oplus su(M) \oplus u(1)$ currents can be constructed as
\begin{align}
\begin{aligned}
	&J_f^\alpha = - (  \varphi^{( \rho \bar \imath )} \varphi^{( \bar \sigma i )} )  (t^\alpha)_{ \rho \bar \sigma} \delta_{\bar \imath i}\, , \quad
	J_f^a =   ( \varphi^{( \rho \bar \imath )}  \varphi^{( \bar \rho j )} )  (t^a)_{j \bar \imath }  \delta_{\rho \bar \rho}  \, , \\
	&J_f^{u(1)} = ( \varphi^{(\rho \bar \imath)} \varphi^{(\bar \rho i)} ) \delta_{\rho \bar \rho} \delta_{\bar \imath i} \, .
\end{aligned}
	\label{Jfs0}
\end{align}
The currents in the denominator of \eqref{superalgebrascoset} are then given by
\begin{align}
	\tilde J^\alpha = J^\alpha + J_f^\alpha \, , \quad
	\tilde J^{u(1)} = \sqrt{MN (M - N)} J^{u(1)} + (N  - M) J_f^{u(1)} \, .
	\label{tJss}
\end{align}

There are two sets of $su(M)$ currents and one $u(1)$ current in the $\mathcal{N}=2$ W-algebra.
One of the sets of $su(M)$ currents is $J^a$ and the other is $K^a = J_f^a$ in \eqref{Jfs0}.
The $u(1)$ current is
\begin{align}
	K = \frac{1}{M - N -  k} \left( \sqrt{M N (M - N)} J^{u(1)} - k J_f^{u(1)} \right) \, .
\end{align}
We have
\begin{align}
	\begin{aligned}
		&G^- = (  \varphi^{(\rho \bar \imath)}  J^{(\bar \rho i)})  \delta_{\rho \bar \rho} \delta_{\bar \imath i} \, , \quad
		G^+ =  (   J^{(\rho \bar \imath)}  \varphi^{(\bar \rho i)} ) \delta_{\rho \bar \rho} \delta_{\bar \imath i} \, , \\
		&G^{-,a} =- (\varphi^{(\rho \bar \imath)} J^{(\bar \rho j)}) (t^a)_{j \bar \imath} \delta_{\rho \bar \rho} \, , \quad G^{+,a} = - (J^{(\rho \bar \imath)} \varphi^{(\bar \rho j)}) (t^a)_{j \bar \imath} \delta_{\rho \bar \rho}
	\end{aligned}
\end{align}
for spin 3/2 currents.
The normalizations are chosen so as to satisfy \eqref{KGOPE}, \eqref{JKG0OPE} and \eqref{JKGaOPE}.
A spin 2 current is the energy-momentum tensor from the coset construction \cite{Goddard:1984vk}.
We find the charged spin 2 currents
\begin{align}
\begin{aligned}
Q^a =&   \left[(J^{(\rho \bar \imath)} J^{(\bar \rho i)}) -  (J^{(\bar \rho i)}J^{(\rho \bar \imath)} )  \right] \delta_{\rho \bar \rho} (t^a)_{i \bar \imath} \\
&- \frac{N }{ M - 2 k} d^{a}_{~bc} (J^bJ^c)  - \frac{2}{k} \sqrt{\frac{N (M - N)}{M}} (J^a J^{u(1)}) 
\end{aligned}
\end{align}
satisfy the required OPEs with spin 1 currents \eqref{A1A2OPE} as well as the primary condition w.r.t.~the Virasoro algebra.
The expression is the same as the bosonic one in \eqref{superQa} due to our convention of $Q^a$. We have checked that the OPEs among generators are reproduced up to null vectors for several explicit examples.


\providecommand{\href}[2]{#2}\begingroup\raggedright\endgroup

\end{document}